\documentclass[reprint,amsmath,amssymb,aps,]{revtex4-2}

\usepackage{graphicx}
\usepackage{dcolumn}
\usepackage{bm}
\usepackage{diagbox}
\usepackage{array}
\usepackage{booktabs}
\usepackage{multirow}
\usepackage{amsmath}
\usepackage{amsfonts}
\usepackage{graphicx}
\usepackage{epsfig}
\usepackage{color}
\usepackage{txfonts}
\usepackage[colorlinks,citecolor=blue]{hyperref}

\begin{document}
\title{Chiral and nonreciprocal single-photon scattering in a chiral-giant-molecule\\ waveguide-QED system}
\author{Juan Zhou}
\affiliation{Key Laboratory of Low-Dimensional Quantum Structures and Quantum Control of Ministry of Education, Key Laboratory for Matter Microstructure and Function of Hunan Province, Department of Physics and Synergetic Innovation Center for Quantum Effects and Applications, Hunan Normal University, Changsha 410081, China}
\author{Xian-Li Yin}
\affiliation{Key Laboratory of Low-Dimensional Quantum Structures and Quantum Control of Ministry of Education, Key Laboratory for Matter Microstructure and Function of Hunan Province, Department of Physics and Synergetic Innovation Center for Quantum Effects and Applications, Hunan Normal University, Changsha 410081, China}
\author{Jie-Qiao Liao}
\email{Corresponding author: jqliao@hunnu.edu.cn}
\affiliation{Key Laboratory of Low-Dimensional Quantum Structures and Quantum Control of Ministry of Education, Key Laboratory for Matter Microstructure and Function of Hunan Province, Department of Physics and Synergetic Innovation Center for Quantum Effects and Applications, Hunan Normal University, Changsha 410081, China}

\begin{abstract}
We study chiral and nonreciprocal single-photon scattering in a chiral-giant-molecule waveguide-QED system. Here, the giant molecule consists of two coupled giant atoms, which interact with two linear waveguides, forming a four-port quantum device. We obtain the exact analytical expressions of the four scattering amplitudes using a real-space method. Under the Markovian limit, we find that the single-photon scattering behavior is determined by the coupling strength between the giant atoms and the waveguides, the coupling strength between the two giant atoms, and the nondipole effect caused by the phase accumulation of photons travelling between the coupling points. It is also found that chiral and nonreciprocal single-photon scattering can be realized by introducing the chiral coupling to break the symmetry in the coupling configuration between the giant molecule and the waveguides. In addition, an ideal chiral emitter-waveguide coupling enables a directional single-photon routing. In the non-Markovian regime, the scattering spectra are characterized by more abundant structures with multiple peaks and dips. In particular, we demonstrate that the non-Markovian retarded effect can induce the nonreciprocal single-photon scattering. Our results have potential applications in the design of optical quantum devices involving giant atoms, which can provide an efficient platform for studying chiral quantum optics.
\end{abstract}

\date{\today}
\maketitle
\narrowtext

\section{Introduction}
The interaction between light and atoms is at the heart of quantum optics~\cite{Tannoudji1992}.
Waveguide quantum electrodynamics (QED)~\cite{Roy2017,Gu2017,Sheremet2013}, addressing the interactions of various atoms with running-wave optical fields, provides a good platform for studying light-matter interaction and quantum optical phenomena. There are a lot of theoretical and experimental works investigating the interaction of few~\cite{Shen2005,Shen2007,Zhou2008,Tudela2011,Zheng2013,Pichler2016,Shi2018} or more~\cite{Ramos2014,Tsoi2008,Liao2015,Ke2019} atoms with one-dimensional (1D) waveguides, where atoms can couple to, and interact via, continuous bosonic modes in 1D waveguides. Several quantum devices such as quantum routers~\cite{Zhou2013,Hoi2011,Shomroni2014} and quantum circulators~\cite{W2017} were predicted in waveguide-QED systems. In addition, photon scatterings in waveguides coupled to other scattering targets such as a nonlinear cavity~\cite{J.-Q. Liao2010} and an optomechanical cavity~\cite{J.-Q. Liao2012,J.-Q. Liao2013} have been studied.

Generally, the size of atoms is much smaller than the wavelength of the interacting photons, and hence the atom can be treated as a point when considering the atom-field interactions, as described by the dipole approximation~\cite{Walls2008}. For example, the typical atomic size is of 0.1 nanometers, but the wavelength of the interacting photons is about several hundred nanometers, thus the field experienced by the atom can be considered as a constant field. Recently, the experimental studies of quantum optics has been expanded to the systems with superconducting artificial atoms~\cite{Wendin2017,You and Nori2011,Blais2021} coupled to surface acoustic waves~\cite{Gustafsson2014,Manenti2017,Andersson2019} or microwave waveguides~\cite{Vadiraj2021,Kannan2020}, where the size of an artificial atom could be on the same order of magnitude as the wavelength of the field and then the dipole approximation no longer holds. At this point, the giant atom~\cite{Kockum2021} as an emerging setup, becomes an effective platform to deal with this problem. In particular, the coupling of giant atoms with running waves becomes an interesting research topic, because the photon/phonon-transport effect needs to be considered in the interaction. Namely, the giant atom will couple to the photons/phonons during a range of interaction region. To simplify the continuous coupling effect and better exhibit the physics, some studies have considered the multiple-point coupling between giant atoms and waveguides~\cite{Kannan2020,Kockum2014,Kockum2018,Cai2021}. The multiple-point coupling leads to a variety of quantum interference effects that are not found in small atoms, such as frequency-dependent Lamb shifts and relaxation rates~\cite{Kockum2014}, interatomic interactions without decoherence~\cite{Kockum2018,Carollo2020,Kannan2020,Cilluffo2020,Soro2022}, and creation of bound states~\cite{Guo2020,Zhao2020,Wang2021}.

The waveguide-QED systems provide an efficient platform for controlling the flow of photons, especially realizing nonreciprocal propagation of photons~\cite{Gonzalez-Ballestero2016,Nie2021,Wang2019,Metelmann2015,Yuan2015,Li2014,Roy2010}, which has wide applications in quantum optical devices. Chiral waveguide-QED systems break the time-reversal symmetry of the systems and make the interactions between atoms and waveguide modes directionally dependent, which can realize quantum information processing tasks that cannot be accomplished by the bidirectional quantum channels ~\cite{sollner2015,Mitsch2014,Petersen2014,Young2015,Grankin2018,Feber2015,Bliokh2015,Lodahl2017}. Recently, the concept of chiral quantum optics has been introduced into giant atomic structures~\cite{Soro2022,Xin Wang2022,Chen2022,Lei Du2021,Carollo2020,Cilluffo2020}, suggesting the possibility of combining the advantages of both paradigms. Moreover, it is an interesting topic to study the non-Markovian effects in chiral optical systems involving giant atoms~\cite{Lei Du2021}. Actually, if the propagating time of photons/phonons between different coupling points is comparable to the atomic lifetime, such effects should be taken into account~\cite{Andersson2019,Guo2020,Lei Du2021,X.-L. Yin2022,L. Du2022,Guo2017}. 

Motivated by these advances, here we study the single-photon scattering in a chiral-giant-molecule waveguide-QED system~\cite{Xian-Li Yin2022}, where the giant molecule consists of two interacting giant atoms. The giant molecule couples to two linear waveguides, forming a four-port device. The exact analytical scattering amplitudes are obtained by using the real-space method, which is valid in both the Markovian and non-Markovian regimes. It is found that the single-photon scattering behavior is strongly influenced by the phase shifts between the coupling points, the coupling strength between two giant atoms, and the coupling strength between giant atoms and waveguides. In the Markovian regime, the phase shifts are detuning-independent. We find that the single-photon scattering is achiral or reciprocal in the symmetric-coupling case, whereas the single-photon scattering is chiral or nonreciprocal in the chiral-coupling case. To realize perfect chirality and nonreciprocity, we further investigate the single-photon scattering in the ideal chiral-coupling case and demonstrate the single photon directional router. In the non-Markovian regime, we show that the scattering spectra exhibit more abundant structures with multiple peaks and staggered dips, which is induced by the detuning-dependent phase shift. In addition, we find that the non-Markovian retarded effect can enhance the nonreciprocity of single-photon scattering in the chiral-coupling case, while this feature does not appear in small-chiral-molecule systems.

The rest of this paper is organized as follows. In Sec.~\ref{Physical model and Eqs}, we introduce the physical model and present the Hamiltonians and scattering solutions. In Sec.~\ref{NACSPscattering of GMWQED}, we analyze single-photon scattering in the cases of symmetric coupling and chiral coupling in the Markovian regime. In Sec.~\ref{Non-Markovian regime}, we study the single-photon scattering in the non-Markovian regime. In Sec.~\ref{Experimental}, we present some discussions on the experimental implementation of this scheme. Finally, we conclude this work in Sec.~\ref{Conclusions}. 
\begin{figure}[tbp]
	\center\includegraphics[width=0.5\textwidth]{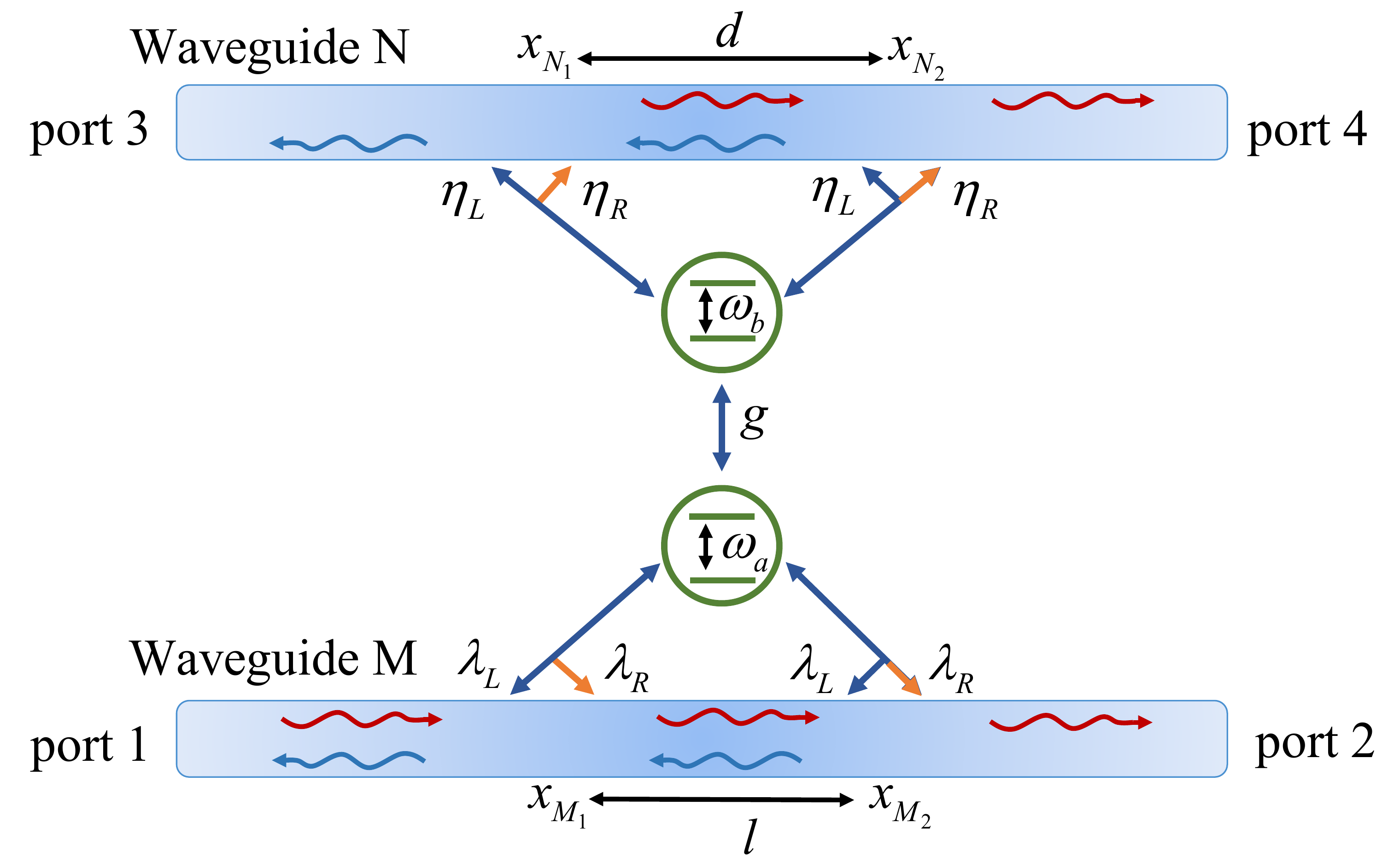}
	\caption{Schematic of a chiral-giant-molecule double-waveguide-QED system, where the giant-molecule consists of two interacting giant atoms $a$ and $b$ with $g$ being the coupling strength. The two-level giant atom $a$ ($b$) chirally couples to waveguide M (N) via two coupling points located at $x_{M_{1}}=0$ and $x_{M_{2}}=l$ ($x_{N_{1}}=0$ and $x_{N_{2}}=d$). }
	\label{model}
\end{figure}

\section{Physical model and Hamiltonians}\label{Physical model and Eqs}
We consider a chiral-giant-molecule waveguide-QED system, in which the giant molecule couples to two waveguides, M and N, as shown in Fig.~\ref{model}. Here, the giant molecule consists of two interacting giant atoms $a$ and $b$ with coupling strength $g$, and each giant atom chirally couples to a corresponding waveguide via two coupling points with a distance ($l$ or $d$). 
The total Hamiltonian of the system consists of three parts $(\hbar=1)$
\begin{equation}
	\hat{H}=\hat{H}_{s}+\hat{H}_{w}+\hat{H}_{I},\label{H}
\end{equation}
where
\begin{subequations}
	\begin{align}
	\hat{H}_{s} =&\omega _{a}\hat{\sigma}_{a}^{+}\hat{\sigma}_{a}^{-}+\omega
	_{b}\hat{\sigma}_{b}^{+}\hat{\sigma}_{b}^{-}, 	\label{Hs}\\
	\hat{H}_{w}=&\sum_{p=M,N}i\upsilon_{g}\int \hat{c}_{Lp}^{\dagger }\left( x\right) 
	\frac{\partial }{\partial x}\hat{c}_{Lp}\left( x\right) dx  \nonumber \\
	&-\sum_{p=M,N}i\upsilon_{g}\int \hat{c}_{Rp}^{\dagger }\left( x\right) \frac{%
		\partial }{\partial x}\hat{c}_{Rp}\left( x\right) dx,	\label{Hw} \\
	\hat{H}_{I} =&\int dxA\left( x\right) \left[ \lambda _{R}\hat{\sigma}%
	_{a}^{+}\hat{c}_{RM}\left( x\right) +\lambda _{L}\hat{\sigma}_{a}^{+}\hat{c}%
	_{LM}\left( x\right) \right]   \nonumber \\
   &+\int dxB\left( x\right) [ \eta _{R}\hat{\sigma}_{b}^{+}\hat{c}%
	_{RN}\left( x\right) +\eta _{L}\hat{\sigma}_{b}^{+}\hat{c}_{LN}\left(
	x\right) ]   \nonumber \\
	&+g\hat{\sigma}_{a}^{-}\hat{\sigma}_{b}^{+}+\text{H.c.},	\label{HI}
\end{align}
\end{subequations}
 with $A\left(x\right)=\delta\left( x\right)+\delta\left(x-l\right)$ and $B\left( x\right) =\delta \left( x\right) +\delta \left( x-d\right)$ [$\delta \left( x\right)$ being the Dirac delta function]. The term $\hat{H}_{s}$ is the free Hamiltonian of the two giant atoms, where $\omega _{a}$ ($\omega _{b}$) is the transition frequency of the two-level giant atom $a$ ($b$), and $\hat{\sigma}_{a}^{+}$ $( \hat{\sigma}%
_{b}^{+}) =\left( \hat{\sigma}_{a}^{-}\right) ^{\dag }$ $[ ( \hat{\sigma}_{b}^{-}) ^{\dag }] =\vert e\rangle_{aa}\langle g\vert $ $\left( \vert e\rangle_{bb}\langle g\vert \right) $ is the raising operator of the atom $a$ ($b$) with the ground state $\left\vert g\right\rangle _{a}$
($\left\vert g\right\rangle _{b}$) and the excited state $\left\vert e\right\rangle _{a}$ ($\left\vert e\right\rangle _{b}$). The term $\hat{H}_{w}$ is the free Hamiltonian of the two waveguides, where   $\hat{c}_{Rp}^{\dagger }\left(
x\right) =[ \hat{c}_{Rp}\left( x\right) ] ^{\dag }$ and $\hat{c}%
_{Lp}^{\dagger }\left( x\right) =[ \hat{c}_{Lp}\left( x\right) ]
^{\dag }$ are the field operators for creating a right- and left-moving photon in waveguide $p$ at position $x$ with the group velocity $\upsilon_{g}$. The last term $\hat{H}_{I}$ describes the interaction between the giant atoms and the waveguides, with $\lambda _{R}$\ and $\lambda _{L}$ ($\eta _{R}$ and $\eta _{L}$) being the coupling strengths between the giant atom $a$ ($b$) and the waveguide M (N). 

The total excitation-number operator of the giant molecule and the waveguides is a conserved quantity. For studying the single-photon scattering, we restrict the system in the single-excitation subspace, then the eigenstate $\left\vert \psi \right\rangle $ of the system can be expressed as
\begin{eqnarray}
	\label{Eigenstate}
	\left\vert \psi \right\rangle  &=&\sum_{p=M,N}\int dx[ \phi _{Rp}\left(
	x\right) \hat{c}_{Rp}^{\dagger }\left( x\right) +\phi _{Lp}\left( x\right) 
	\hat{c}_{Lp}^{\dagger }\left( x\right) ] \left\vert \emptyset
	\right\rangle   \nonumber \\
	&&+u_{a}\hat\sigma _{a}^{+}\left\vert \emptyset \right\rangle +u _{b}\hat\sigma
	_{b}^{+}\left\vert \emptyset \right\rangle,
\end{eqnarray}
where $\phi _{Rp}\left( x\right) $ and $ \phi _{Lp}\left( x\right) $ are the single-photon wave functions of the right- and left-propagating photon at position $x$ in the waveguide $p$. The coefficients $u_{a}$ and $u_{b}$ are the corresponding excitation probability amplitudes of giant atoms $a$ and $b$, respectively. The state $\left\vert \emptyset \right\rangle$  represents the vacuum state, which means that there are no photons in the waveguides and both giant atoms $a$ and $b$ are in their ground states. Based on the stationary Schr\"{o}dinger equation $\hat{H}\left\vert \psi \right\rangle =E\left\vert \psi \right\rangle $, the probability amplitudes are determined by the following equations:
\begin{eqnarray}
	E\phi _{RM}\left( x\right)  &=&-i\upsilon_{g}\frac{\partial }{\partial x}\phi
	_{RM}\left( x\right) +u_{a}\lambda _{R}A\left( x\right) ,  \nonumber \\
	E\phi _{LM}\left( x\right)  &=&i\upsilon_{g}\frac{\partial }{\partial x}\phi
	_{LM}\left( x\right) +u_{a}\lambda _{L}A\left( x\right) ,  \nonumber \\
	E\phi _{RN}\left( x\right)  &=&-i\upsilon_{g}\frac{\partial }{\partial x}\phi
	_{RN}\left( x\right) +u_{b}\eta _{R}B\left( x\right) ,  \nonumber \\
	E\phi _{LN}\left( x\right)  &=&i\upsilon_{g}\frac{\partial }{\partial x}\phi
	_{LN}\left( x\right) +u_{b}\eta _{L}B\left( x\right) ,\label{Equations of motion1 }
\end{eqnarray}
and
\begin{eqnarray}
	\Delta _{a} u _{a} &=&g u _{b}+\lambda _{R}\left[ \phi _{RM}\left(
	0\right) +\phi _{RM}\left( l\right) \right]   \nonumber \\
	&&+\lambda _{L}\left[ \phi _{LM}\left( 0\right) +\phi _{LM}\left( l\right) %
	\right] ,  \nonumber \\
	\Delta _{b} u _{b} &=&g u _{a}+\eta _{R}\left[ \phi _{RN}\left( 0\right)
	+\phi _{RN}\left( d\right) \right]   \nonumber \\
	&&+\eta _{L}\left[ \phi _{LN}\left( 0\right) +\phi _{LN}\left( d\right) %
	\right] ,\label{Equations of motion2 }
\end{eqnarray}	
where $\Delta _{a}=E-\omega _{a}$ ($\Delta _{b}=E-\omega _{b}$) is the frequency detuning between the propagating photon in waveguide M (N) and the atomic transition $\left\vert g\right\rangle _{a}\leftrightarrow \left\vert e\right\rangle _{a}$ $\left( \left\vert g\right\rangle _{b}\leftrightarrow \left\vert
e\right\rangle _{b}\right) $. 

We first consider the case where a single photon with energy $E$ is injected from the left-hand side of waveguide M (port 1). Then the wave functions $\phi _{RM}\left( x\right) ,\phi _{LM}\left( x\right) ,\phi _{RN}\left(x\right) $, and $\phi _{LN}\left( x\right) $ can be written as
\begin{eqnarray}
\phi _{RM}\left( x\right)  &=&e^{ik_{m}x}\left[ \theta \left( -x\right)
+t_{l}\theta \left( x\right) \theta \left( l-x\right) +t_{M}\theta \left(
x-l\right) \right] ,  \nonumber \\
\phi _{LM}\left( x\right)  &=&e^{-ik_{m}x}\left[ r_{M}\theta \left(
-x\right) +r_{l}\theta \left( x\right) \theta \left( l-x\right) \right] , 
\nonumber \\
\phi _{RN}\left( x\right)  &=&e^{ik_{n}x}\left[ t_{d}\theta \left( x\right)
\theta \left( d-x\right) +t_{N}\theta \left( x-d\right) \right] ,  \nonumber
\\
\phi _{LN}\left( x\right)  &=&e^{-ik_{n}x}\left[ r_{N}\theta \left(
-x\right) +r_{d}\theta \left( x\right) \theta \left( d-x\right) \right] ,\label{WavefunctionL}
\end{eqnarray}
with the Heaviside step function $\theta \left(x\right)$, which is used to distinguish different intervals. Here $e^{ik_{m}x}\theta \left( -x\right) $ represents that the single photon is injected from port 1. The function $t_{l}e^{ik_{m}x}\theta \left( x\right) \theta \left(
l-x\right) $ $[ t_{d}e^{ik_{n}x}\theta \left( x\right) \theta \left(
d-x\right) ] $ is the wave function of the single right-propagating photon between $x=0$ and $x=l$ $\left( x=d\right) $ in waveguide M (N) with the
transmission amplitude $t_{l}$ $\left( t_{d}\right) $. The $r_{l}e^{-ik_{m}x}\theta
\left( x\right) \theta \left( l-x\right) $ $[ r_{d}e^{-ik_{n}x}\theta
\left( x\right) \theta \left( d-x\right) ] $ is the wave function of
the single left-propagating photon between $x=0$ and $x=l$ $\left(
x=d\right) $ in waveguide M (N) with the reflection
amplitude $r_{l}$ $\left( r_{d}\right) $. $t_{M}$ and $r_{M}$ $\left( t_{N}\text{ and }r_{N}\right) $ are the transmission and reflection amplitudes of the input photon in waveguide M (N),
respectively. The single right-propagating photon accumulates a phase term $e^{ik_{m}l}$ ($e^{ik_{n}d}$) when it passes through the two coupling points of waveguide M (N) with the giant atom $a$ ($b$). Correspondingly, the photon accumulates a conjugate phase term $e^{-ik_{m}l}$ ($e^{-ik_{n}d}$) during the propagation in the opposite direction. 

To study chiral and nonreciprocal scattering properties of the system, we then consider that a single photon with energy $E$ is injected from the right-hand side of waveguide M (port 2). The wave functions $\tilde{\phi}_{RM}\left( x\right) ,\tilde{\phi}_{LM}\left( x\right) ,\tilde{\phi}_{RN}\left( x\right) $, and $\tilde{\phi}_{LN}\left( x\right) $ in this case can be written as
\begin{eqnarray}
	\tilde{\phi}_{RM}\left( x\right)  &=&e^{ik_{m}x}\left[ \tilde{r}_{l}\theta
	\left( x\right) \theta \left( l-x\right) +\tilde{r}_{M}\theta \left(
	x-l\right) \right] ,  \nonumber \\
	\tilde{\phi}_{LM}\left( x\right)  &=&e^{-ik_{m}x}\left[ \tilde{t}_{M}\theta
	\left( -x\right) +\tilde{t}_{l}\theta \left( x\right) \theta \left(
	l-x\right) +\theta \left( x-l\right) \right] ,  \nonumber \\
	\tilde{\phi}_{RN}\left( x\right)  &=&e^{ik_{n}x}\left[ \tilde{r}_{d}\theta
	\left( x\right) \theta \left( d-x\right) +\tilde{r}_{N}\theta \left(
	x-d\right) \right] ,  \nonumber \\
	\tilde{\phi}_{LN}\left( x\right)  &=&e^{-ik_{n}x}\left[ \tilde{t}_{N}\theta
	\left( -x\right) +\tilde{t}_{d}\theta\left(x\right) \theta \left(
	d-x\right) \right],\label{WavefunctionR}
\end{eqnarray}
where $e^{-ik_{m}x}\theta \left( x-l\right) $ represents that the single photon is injected from port 2. The $\tilde{t}_{M}$ ($\tilde{t}_{N}$) and $\tilde{r}_{M}$ ($\tilde{r}_{N}$) denote that the single photon transmitted out or reflected back in waveguide M (N), respectively, when the single photon is injected from port 2 of waveguide M.

\section{Single-photon scattering in the Markovian regime}\label{NACSPscattering of GMWQED}
In this section, we study the single-photon scattering in the chiral-giant-molecule two-waveguide system in the Markovian regime. To this end, we first calculate the single-photon transmission and reflection amplitudes and then obtain the corresponding transmission and reflection coefficients.

\subsection{The transmission and reflection coefficients}\label{The transmission reflection coefficients}
We begin by considering the case where a single photon is injected from port 1 of waveguide M. Substitution of Eq.~(\ref{WavefunctionL}) into Eqs.~(\ref{Equations of motion1 }) and (\ref{Equations of motion2 }) yields the following relations:
\begin{eqnarray}
	i\upsilon_{g}\left( 1-t_{l}\right) +u _{a}\lambda _{R} &=&0,  \nonumber \\
	i\upsilon_{g}\left( r_{l}-r_{M}\right) +u _{a}\lambda _{L} &=&0,  \nonumber \\
	-i\upsilon_{g}t_{d}+u _{b}\eta _{R} &=&0,  \nonumber \\
	i\upsilon_{g}\left( r_{d}-r_{N}\right) +u_{b}\eta _{L} &=&0,  \nonumber \\
	i\upsilon_{g}\left( t_{l}-t_{M}\right) e^{ik_{m}l}+u _{a}\lambda _{R} &=&0, 
	\nonumber \\
	-i\upsilon_{g}r_{l}e^{-ik_{m}l}+u _{a}\lambda _{L} &=&0,  \nonumber \\
	i\upsilon_{g}\left( t_{d}-t_{N}\right) e^{ik_{n}d}+u _{b}\eta _{R} &=&0, 
	\nonumber \\
	-i\upsilon_{g}r_{d}e^{-ik_{n}d}+u _{b}\eta _{L} &=&0,\label{Eqs of motion1}
\end{eqnarray}
and
\begin{eqnarray}
	\Delta u_{a} &=&g u _{b}+\frac{1}{2}\left[ \lambda _{R}\left(
	1+t_{l}\right) +\lambda _{L}\left( r_{M}+r_{l}\right) \right]   \nonumber \\
	&&+\frac{1}{2}[ \lambda _{R}\left( t_{l}+t_{M}\right)
	e^{ik_{m}l}+\lambda _{L}r_{l}e^{-ik_{m}l}] ,  \nonumber \\
	\Delta u_{b} &=&gu_{a}+\frac{1}{2}\left[ \eta _{R}t_{d}+\eta
	_{L}\left( r_{N}+r_{d}\right) \right]   \nonumber \\
	&&+\frac{1}{2}[ \eta _{R}\left( t_{d}+t_{N}\right) e^{ik_{n}d}+\eta
	_{L}r_{d}e^{-ik_{n}d}].\label{Eqs of motion2}
\end{eqnarray}
In Eq.~(\ref{Eqs of motion2}), we assumed that the two giant atoms have the same transition frequency, i.e., $\Delta =\Delta _{a}=\Delta _{b}=E-\omega _{0}$. By solving Eqs.~(\ref{Eqs of motion1}) and (\ref{Eqs of motion2}), the scattering amplitudes can be obtained as
	\begin{subequations}
		\label{Sca- amplitude port1}
		\begin{align}
		r_{N} &=\frac{i( 1+e^{i\varphi _{d}}) ( 1+e^{i\varphi
				_{l}}) g\sqrt{\Gamma _{\eta L}\Gamma _{\lambda R}}}{g^{2}-(
			\Delta +F_{\eta }) ( \Delta +F_{\lambda }) }, \label{Sca- amplitude port1 1}\\
		t_{N} &=\frac{i( 1+e^{-i\varphi _{d}}) ( 1+e^{i\varphi
				_{l}}) g\sqrt{\Gamma _{\eta R}\Gamma _{\lambda R}}}{g^{2}-(
			\Delta +F_{\eta }) ( \Delta +F_{\lambda }) }, \label{Sca- amplitude port1 2} \\
		r_{M} &=\frac{i( 1+e^{i\varphi _{l}}) ^{2}( \Delta +F_{\eta
			}) \sqrt{\Gamma _{\lambda L}\Gamma _{\lambda R}}}{g^{2}-( \Delta
			+F_{\eta }) ( \Delta +F_{\lambda }) },  \label{Sca- amplitude port1 3}\\
		t_{M} &=\frac{g^{2}-[ i( 1+e^{i\varphi _{l}}) ( \Gamma
			_{\lambda L}-e^{-i\varphi _{l}}\Gamma _{\lambda R})+\Delta ]
			( F_{\eta }+\Delta) }{g^{2}-( \Delta +F_{\eta })
		( \Delta +F_{\lambda }) }, \label{Sca- amplitude port1 4}
		\end{align}
	\end{subequations}
with
\begin{subequations}
	\begin{align}
F_{\lambda } &=i( 1+e^{i\varphi _{l}}) \left( \Gamma _{\lambda
	L}+\Gamma _{\lambda R}\right) , \\
F_{\eta } &=i( 1+e^{i\varphi _{d}}) ( \Gamma _{\eta L}+\Gamma
_{\eta R}) . 
\end{align}
\end{subequations}

 By defining the scattering coefficients $R_{N}=\left\vert r_{N}\right\vert ^{2}$, $T_{N}=\left\vert t_{N}\right\vert^{2}$, $R_{M}=\left\vert r_{M}\right\vert ^{2}$, and $T_{M}=\left\vert t_{M}\right\vert ^{2}$, we can prove $R_{N}+T_{N}+R_{M}+T_{M}=1$ due to the energy conservation. In Eq.~(\ref{Sca- amplitude port1}), the variables $\Gamma _{\lambda L}=\lambda _{L}^{2}/\upsilon_{g}$ and $\Gamma _{\lambda
 	R}=\lambda _{R}^{2}/\upsilon_{g}$ ($\Gamma _{\eta L}=\eta _{L}^{2}/\upsilon_{g}$ and $\Gamma
 _{\eta R}=\eta _{R}^{2}/\upsilon_{g}$) are the decay rates from the excited state 
 $\left\vert e\right\rangle_{a}$ ($\left\vert e\right\rangle_{b}$)
 to the ground state $\left\vert g\right\rangle _{a}$ ($\left\vert g\right\rangle _{b}$) induced by the left- and right-propagating waveguide modes, respectively. The $\varphi _{l}=lk_{m}$ ($\varphi _{d}=dk_{n}$) is the accumulated phase when the single photon passes through the two coupling points of giant atom $a$ ($b$) with waveguide M (N).  According to the relations $\Delta =E-\omega _{0}$ and $E=\upsilon_{g}k_{p}$, both $\varphi _{l}$ and $\varphi_{d}$ can be written as a $\Delta $-dependent part plus a constant part: $\varphi _{l}=\tau _{l}\Delta+\theta _{l}$ and $\varphi _{d}=\tau _{d}\Delta +\theta _{d}$, with $\tau_{l}=l/\upsilon_{g}$ ($\tau _{d}=d/\upsilon_{g}$) being the propagating time of the single photon between the two coupling points of giant atom $a$ ($b$). We point out that Eq.~(\ref{Sca- amplitude port1}) is valid in both the Markovian and non-Markovian regimes. Therefore, we can study the single-photon scattering in these two regimes, depending on whether
 	the propagation time $\tau _{l}$ and $\tau _{d}$ can be neglected. In this section, we focus on single-photon scattering in the Markovian regime, in
 	which the propagation time $\tau _{l}$ and $\tau _{d}$ are much less than the life times of giant atoms, i.e., $\{ \tau _{l}\Gamma _{\lambda R},\tau _{l}\Gamma _{\lambda L}\}
 	\ll 1$ and $\{ \tau _{d}\Gamma _{\eta R},\tau _{d}\Gamma _{\eta L}\} \ll 1$, and then the accumulated phase shifts can be approximated as $\varphi_{l}\approx \theta _{l}$ and $\varphi _{d}\approx \theta  _{d}$.

Similarly, for a single photon injected from port 2 of waveguide M, we can obtain the scattering amplitudes as
	\begin{subequations}\label{Sca- amplitude port2}
		\begin{align}
			\tilde{r}_{N} &=\frac{i( 1+e^{-i\varphi _{d}}) (
				1+e^{-i\varphi _{l}}) g\sqrt{\Gamma _{\eta R}\Gamma _{\lambda L}}}{%
				g^{2}-( \Delta +F_{\eta }) ( \Delta +F_{\lambda }) },\label{Sca- amplitude port2 1}
			\\
			\tilde{t}_{N} &=\frac{i( 1+e^{i\varphi _{d}}) (
				1+e^{-i\varphi _{l}}) g\sqrt{\Gamma _{\eta L}\Gamma _{\lambda L}}}{%
				g^{2}-( \Delta +F_{\eta }) \left( \Delta +F_{\lambda }\right) },\label{Sca- amplitude port2 2}
			\\
			\tilde{r}_{M} &=\frac{i( 1+e^{-i\varphi _{l}}) ^{2}( \Delta
				+F_{\eta }) \sqrt{\Gamma _{\lambda L}\Gamma _{\lambda R}}}{%
				g^{2}-( \Delta +F_{\eta }) \left( \Delta +F_{\lambda }\right) },\label{Sca- amplitude port2 3}
			\\
			\tilde{t}_{M} &=\frac{g^{2}-[ i( 1+e^{i\varphi _{l}})(
				\Gamma _{\lambda R}-e^{-i\varphi _{l}}\Gamma _{\lambda L}) +\Delta %
			] ( F_{\eta }+\Delta ) }{g^{2}-( \Delta +F_{\eta
				})( \Delta +F_{\lambda }) }.\label{Sca- amplitude port2 4}
		\end{align}
	\end{subequations}
Here, the scattering coefficients are defined as $\tilde{R}_{N}=\vert\tilde{r} _{N}\vert ^{2}$, $\tilde{T}_{N}=\vert \tilde{t}_{N}\vert^{2}$, $\tilde{R}_{M}=\vert \tilde{r}_{M}\vert ^{2}$, and $\tilde{T}_{M}=\vert \tilde{t}_{M}\vert ^{2}$.
It can be seen from Eqs.~(\ref{Sca- amplitude port1}) and (\ref{Sca- amplitude port2}) that when considering a single-photon injection from port 1 and port 2, respectively, the transmission of the photon from port 1 (2) to other three ports are nonreciprocal under the condition $\Gamma _{\eta R}\neq \Gamma _{\eta L}$ or (and) $\Gamma _{\lambda R}\neq \Gamma _{\lambda L}$, and that the reflection in waveguide M is always reciprocal. We would like to point out that the scattering amplitudes for the single-photon injection from ports 3 and 4 can also be obtained via the same method. Therefore, the scattering matrix is defined by $S=[s_{ij}]$ (for $i$, $j=1\text{-}4$) with $s_{ij}$ being the scattering amplitudes from ports $j$ to $i$. It has been shown that when the off-diagonal elements $s_{ij}$ (for $i\neq j$) in $S$ satisfy the relation  $s_{ij}\neq s_{ji}$, the single photon transmission between the two ports (from ports $j$ to  $i$ and from ports $i$ to $j$) is nonreciprocal~\cite{Caloz2018}. In this work, we only focus on the single-photon scattering from ports 1 and 2. To better understand the nonreciprocal scattering, we will discuss in detail the scattering behaviors of this chiral-giant-molecule waveguide-QED system in both the symmetric- and chiral-coupling cases.

\subsection{Symmetric-coupling case}\label{Symmetric coupling case}
In this section, we study single-photon scattering in the symmetric-coupling case $\lambda _{R}=\lambda _{L}=\eta _{R}=\eta _{L}$ ($\Gamma _{R}=\Gamma _{L}=\Gamma _{R}=\Gamma _{L}=\Gamma$). By substituting the symmetry conditions into Eqs.~(\ref{Sca- amplitude port1}) and (\ref{Sca- amplitude port2}), the scattering amplitudes are reduced to
\begin{subequations}\label{Sca- amplitude1}
	\begin{align}
		r_{N} =&e^{-i\theta _{d}}e^{-i\theta _{l}}\tilde{r}_{N}=\frac{i( 1+e^{i\theta
				_{d}}) ( 1+e^{i\theta _{l}}) g\Gamma }{g^{2}-\left( \Delta
			+F_{d}\right) \left( \Delta +F_{l}\right) }, \\
		t_{N} =&e^{i\theta _{d}}e^{-i\theta _{l}}\tilde{t}_{N}=\frac{i(
			1+e^{-i\theta _{d}}) ( 1+e^{i\theta _{l}}) g\Gamma }{%
			g^{2}-\left( \Delta +F_{d}\right) \left( \Delta +F_{l}\right) }, \\
		r_{M} =&e^{-2i\theta _{l}}\tilde{r}_{M}=\frac{i( 1+e^{i\theta _{l}})
			^{2}\left( \Delta +F_{d}\right) \Gamma }{g^{2}-\left( \Delta +F_{d}\right)
			\left( \Delta +F_{l}\right) }, \\
		t_{M} =&\frac{\ \ g^{2}-\left( \Delta -2\Gamma \sin \theta _{l}\right) \left(
			\Delta +F_{d}\right) }{g^{2}-\left( \Delta +F_{d}\right) \left( \Delta
			+F_{l}\right) },
	\end{align}
\end{subequations}
with
\begin{subequations}
	\begin{align}
		F_{l} =&2i( 1+e^{i\theta _{l}}) \Gamma , \\
		F_{d} =&2i( 1+e^{i\theta _{d}}) \Gamma .		
	\end{align}
\end{subequations}
Equation~(\ref{Sca- amplitude1}) shows that the scattering coefficients satisfy the relations $T_{N}=\tilde{T}_{N}$, $R_{N}=\tilde{R}_{N}$, $%
T_{M}=\tilde{T}_{M}$, and $R_{M}=\tilde{R}_{M}$, which indicate that the scattering behavior of the single photon is symmetric. Hence, in the case, we only focus on the scattering behavior for a single photon injected from port 1 of waveguide  M.

\begin{figure}[tbp]
	\center\includegraphics[width=0.5\textwidth]{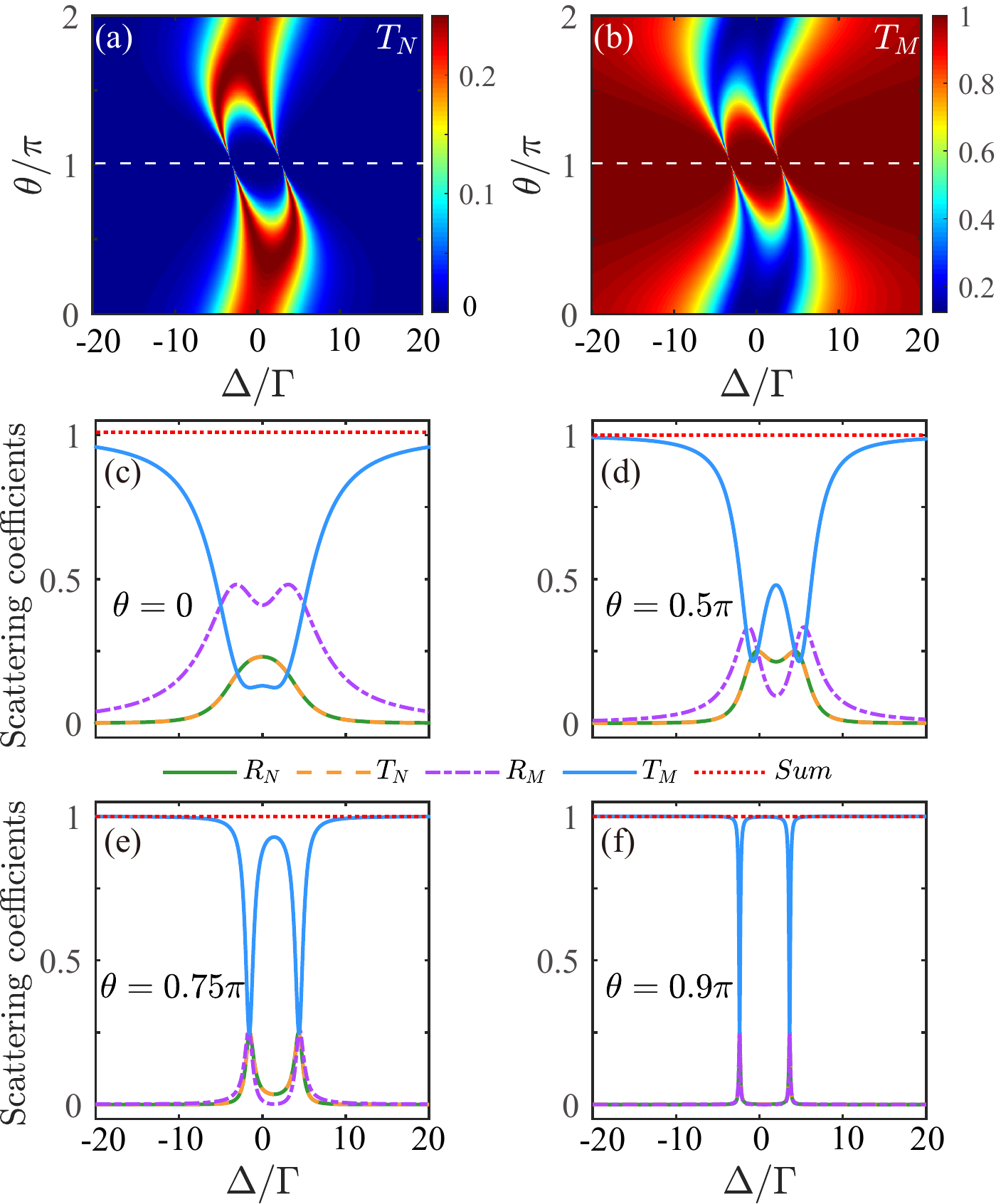}
	\caption{Transmission coefficients (a) $T_{N}$ and (b) $T_{M}$ versus the scaled detuning $\Delta/\Gamma $ and the scaled phase shift $\theta/\pi$. The white dashed lines are used to label the transmission coefficient when $\theta =\pi$.
		 Scattering coefficients $R_{N}$ (green solid curve), $T_{N}$ (yellow dashed curve), $R_{M}$ (purple dot-dashed curve) and $T_{M}$ (blue solid curve) versus the scaled detuning $\Delta/\Gamma $ in the symmetric-coupling case, with (c) $\theta=0$, (d) $\theta=0.5\pi $, (e) $\theta=0.75\pi $, and (f) $\theta=0.9\pi $. The relation $R_{N}+T_{N}+R_{M}+T_{M}=1$ (red dotted curve) is always valid in the above cases. In all panels, we choose $g/\Gamma =3$. }
	\label{Symic-case-TR-vs-Delta-and-Theta}
\end{figure}

\begin{figure}[tbp]
	\center\includegraphics[width=0.49\textwidth]{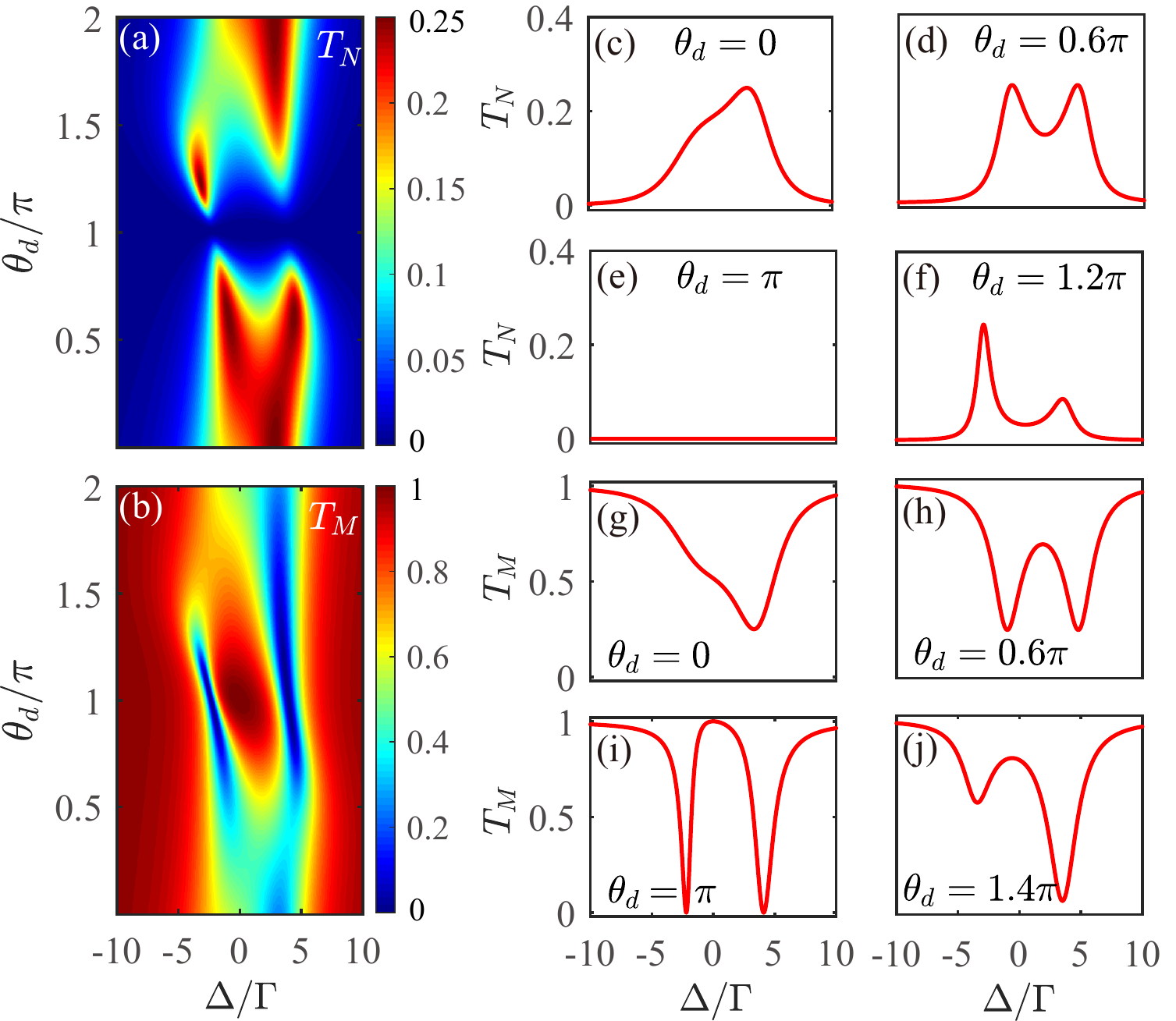}
	\caption{(a) Scattering coefficient $T_{N}$ versus the scaled detuning $\Delta/\Gamma $ and the scaled
		phase shift $\theta _{d}/\pi$. The profiles of panel (a) are shown by the curves
		in panels (c)-(f) at different phases: (c) $\theta _{d}=0$, (d) $\theta _{d}=0.6\pi 
		$, (e) $\theta _{d}=\pi $, and (f) $\theta _{d}=1.2\pi $. (b) Scattering
		coefficient $T_{M}$ versus the scaled detuning $\Delta/\Gamma $ and the scaled phase shift $%
		\theta _{d}/\pi$. The profiles of panel (b) are shown by the curves in (g)-(j)
		at different phases: (g) $\theta _{d}=0$, (h) $\theta _{d}=0.6\pi $, (i) $%
		\theta _{d}=\pi $, and (j) $\theta _{d}=1.4\pi $. In all panels, $g/\Gamma =3
		$ and $\theta _{l}=0.6\pi $.}
	\label{Symic-case-TNTM-vs-Delta-and-Thetad}
\end{figure}

Figure~\ref{Symic-case-TR-vs-Delta-and-Theta}(a) [\ref{Symic-case-TR-vs-Delta-and-Theta}(b)] shows the dependence of the transmission coefficient $T_{N}$ ($T_{M}$) on the scaled detuning $\Delta/\Gamma$ and the scaled phase shift $\theta/\pi$ in the case $\theta_{l}=\theta_{d}=\theta$. Here we find that the period of the two transmission spectra is $2\pi$. To see more about the effect of the phase shifts on the single-photon scattering, in Figs.~\ref{Symic-case-TR-vs-Delta-and-Theta}(c)-\ref{Symic-case-TR-vs-Delta-and-Theta}(f), we show the four scattering coefficients as a function of $\Delta/\Gamma$ for different phase-shift values in the region of $\theta \in \left[ 0,\pi \right] $, due to the relations of $R_{N/M}\left( \Delta ,\theta \right) =R_{N/M}\left( -\Delta ,2\pi -\theta
 \right) $ and $T_{N/M}\left( \Delta ,\theta \right) =R_{N/M}\left( -\Delta ,2\pi -\theta\right) $. One can see that all the scattering spectra are symmetric about $\Delta_{1} =2\Gamma \sin \theta $.
 With the increase of $\theta $, two peaks (or valleys) appear in the scattering spectra, and the position of the peaks (or valleys) changes accordingly. The depth of the peaks (or valleys) increases, and the width of the peaks (or valleys) decreases. The reflection spectrum of $R_{N}$ (green solid curve) coincides with the transmission spectrum of $T_{N}$ (yellow dashed curve). This means that the  probabilities for transmitting the single photon to port $3$ and port $4$ of waveguide N have equal value, i.e., $R_{N}=T_{N}$. When $\theta=0$, $T_{N}$ and $R_{N}$ have only one peak located at $\Delta_{1} =2\Gamma \sin \theta $. The features of $R_{M}$ (purple dot-dashed curve) are different, i.e., there are two peaks at $\theta=0$, and the peak values of the $R_{M}$ are larger than these of $T_{N}$ and $R_{N}$. As $\theta $ increases, the peak values of $R_{M}$ gradually decrease to be comparable to $T_{N}$ and $R_{N}$. From Fig.~\ref{Symic-case-TR-vs-Delta-and-Theta}(e), we notice that when $\theta =0.75\pi $, all scattering coefficients share a common value of almost 0.25 at $\Delta =\Delta _{1}\pm {ie^{-i\theta }\sqrt{e^{2i\theta }\left[
 		-g^{2}+6\Gamma ^{2}+2\Gamma ^{2}\left( 4\cos \theta +\cos \left( 2\theta\right) \right) \right] }}$, which indicates an equal probability of the single photon appearing at all four ports. 
 	
The white dashed lines in Figs.~\ref{Symic-case-TR-vs-Delta-and-Theta}(a) and~\ref{Symic-case-TR-vs-Delta-and-Theta}(b) are used to label the transmission coefficient when $\theta =\pi$. It can be seen that the input single photon is completely transmitted from ports 1 to 2 of waveguide M, which is caused by the decoupling between the giant molecule and the waveguides. However, when $\theta$ approaches to $\left( 2n+1\right) \pi$ with an integer $n$ [for example $\theta=0.9\pi$ in Fig.~\ref{Symic-case-TR-vs-Delta-and-Theta}(f)], the four scattering spectra exhibit Rabi splitting-like line shapes, which are similar to the vacuum Rabi splitting in cavity QED~\cite{Shen2005}. Meanwhile, it can be seen that the peaks of the four spectra have the same locations. To explain this phenomenon, we choose $\theta=\pi+\delta$ with $\left\vert \delta \right\vert \ll 1$ and take the scattering amplitude $t_{N}$ as an example. In this case, $t_{N}$ can be approximated to
\begin{eqnarray}\label{Sca- amplitude4}
t_{N}\approx \frac{ig\Gamma \delta ^{2}}{g^{2}-\left( \Delta +2\Gamma \delta
+i\Gamma \delta ^{2}\right) ^{2}}.
\end{eqnarray}   
According to Eq.~(\ref{Sca- amplitude4}), we obtain the positions of the two splitting peaks of $T_{N}$, which are located at $\Delta =-2\Gamma \delta \pm \sqrt{g^{2}-\Gamma ^{2}\delta ^{4}}$. Then the distance between the two peaks can be calculated as $d_{N}=\left\vert 2\sqrt{g^{2}-\Gamma ^{2}\delta ^{4}}\right\vert $.  Since we consider the case of $\left\vert \delta \right\vert \ll 1$, the distance can be approximated as $d_{N}=2g$. This means that when the phase shift $\theta$ approaches $\pi$, the distance $d_{N}$ only depends on the coupling strength $g$.	

Next, we consider the case of two different phase shifts $\theta _{l}\neq\theta _{d}$. Based on the relations $T_{N}=R_{N}$ and $T_{N}+R_{N}+$ $T_{M}+R_{M}=1$, in the following discussions we focus on the two transmission coefficients $T_{N}$ and $T_{M}$. 
Figures~\ref{Symic-case-TNTM-vs-Delta-and-Thetad}(a) and~\ref{Symic-case-TNTM-vs-Delta-and-Thetad}(b) depict the coefficients $T_{N}$ and $T_{M}$ versus
the scaled detuning $\Delta/\Gamma $\ and the scaled phase shift $\theta _{d}/\pi$, with a given
phase $\theta _{l}=0.6\pi $. Here we find that the transmission spectra $T_{N}$ and $T_{M}$ are asymmetric, which is different from the case of $\theta _{l}= \theta _{d}$. 
From the figures, we know that the line shapes of the transmission spectra exhibit different features at various $\theta _{d}$. To see these features more clearly, the profiles of Figs.~\ref{Symic-case-TNTM-vs-Delta-and-Thetad}(a) and~\ref{Symic-case-TNTM-vs-Delta-and-Thetad}(b) are shown
by the curves in Figs.~\ref{Symic-case-TNTM-vs-Delta-and-Thetad}(c)-\ref{Symic-case-TNTM-vs-Delta-and-Thetad}(j) at different phases $\theta _{d}\in \left[ 0,2\pi \right] $. We can see that the line shapes are non-Lorentzian and the
spectra are symmetric when $\theta _{l}= \theta _{d}$. It is pointed out that $\theta _{d}=0$ corresponds to the case where the giant molecule is coupled to waveguide N at one point. In this case, the transmission peaks in
Fig.~\ref{Symic-case-TNTM-vs-Delta-and-Thetad}(c) and the transmission valleys in Fig.~\ref{Symic-case-TNTM-vs-Delta-and-Thetad}(g) are irregular. As $%
\theta _{d}$ changes, both the size and shape of the transmission peaks and valleys change. For example, in Fig.~\ref{Symic-case-TNTM-vs-Delta-and-Thetad}(f) for $\theta _{d}=1.2\pi $, the transmission peak on the left is higher than the peak on the
right, while in Fig.~\ref{Symic-case-TNTM-vs-Delta-and-Thetad}(j) for $\theta _{d}=1.4\pi $, the transmission valley on the
right is deeper than the one on the left. When $\theta
_{d}=0.6\pi $, the transmission peaks in Fig.~\ref{Symic-case-TNTM-vs-Delta-and-Thetad}(d) and the
transmission valleys in Fig.~\ref{Symic-case-TNTM-vs-Delta-and-Thetad}(h) are symmetric. In addition, when $\theta _{d}=\left( 2n+1\right) \pi $ with an
integer $n$, the transmission peaks in Fig.~\ref{Symic-case-TNTM-vs-Delta-and-Thetad}(e) disappear
completely, which means that the incident photon cannot be
transmitted from waveguides M to N. This is because the giant molecule decouples from waveguide N. 
The above discussion indicates that the phase shifts between the coupling points of the giant atoms affect the scattering of the incident photon in the waveguide.

\begin{figure}[tbp]
	\center\includegraphics[width=0.48\textwidth]{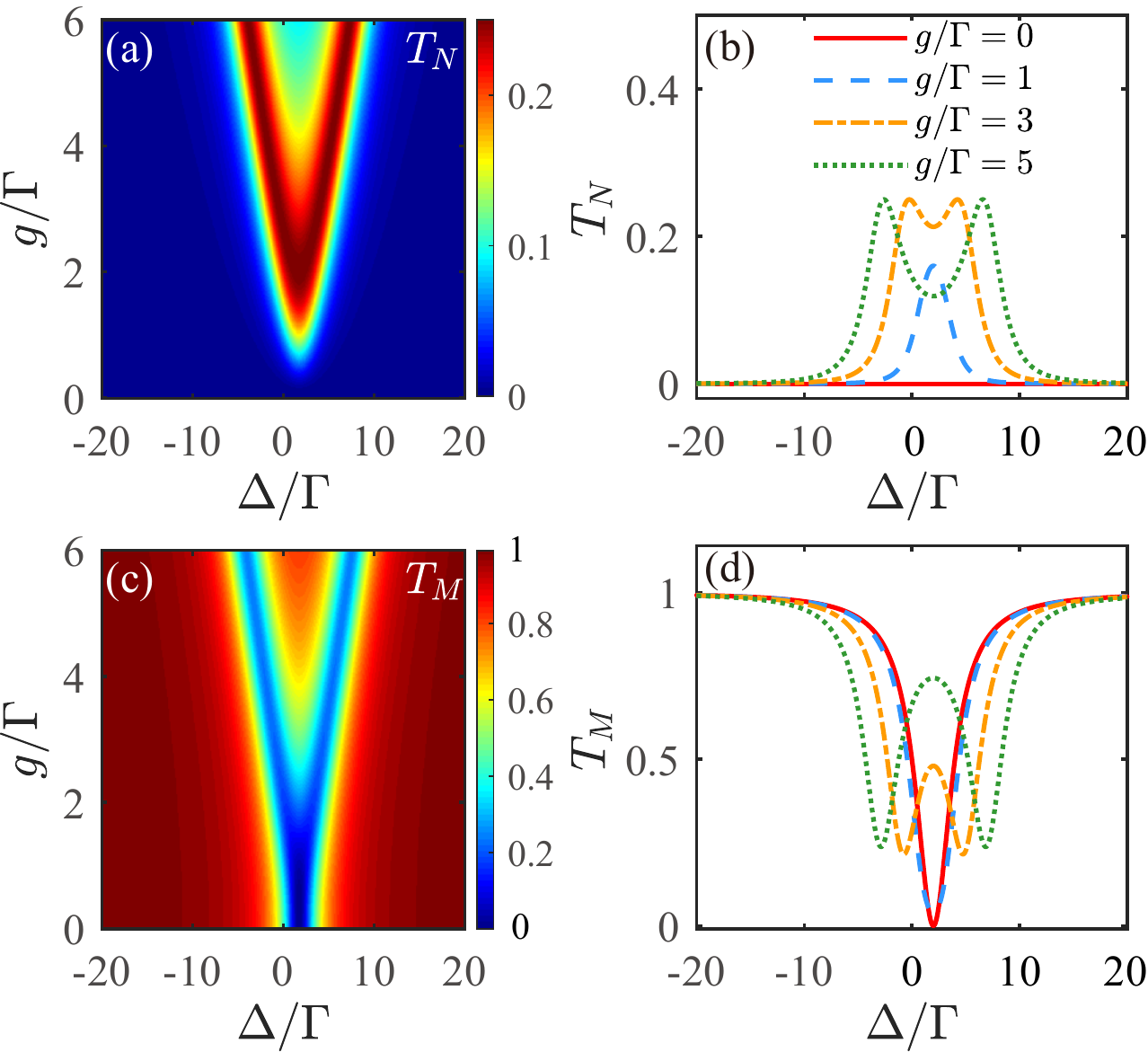}
	\caption{Scattering coefficients (a) $T_{N}$ and (c) $T_{M}$
		as functions of $\Delta/\Gamma $ and $g/\Gamma$. The curves in (b) and (d)
		show the profiles of panel (a) and (c) at $g/\Gamma=0$ (red solid curve), $g/\Gamma=1$ (blue dashed curve), $g/\Gamma=3$ (yellow dot-dashed curve), and $g/\Gamma=5$ (green dotted curve).
		In all panels, we choose $\theta _{l}=\theta _{d}=\theta =0.5\pi $.}
	\label{Symic-case-TR-vs-Delta-and-g}
\end{figure}

In addition to the effect of the phase factors on the single-photon scattering, we are also interested in the effect of the coupling strength between the two giant atoms on the single-photon scattering. We plot the transmission coefficients $T_{N}$ and $T_{M}$ as functions of $\Delta/\Gamma $ and $g/\Gamma$ with $\theta _{l}=\theta _{d}=\theta=0.5\pi$ in Figs.~\ref{Symic-case-TR-vs-Delta-and-g}(a) and \ref{Symic-case-TR-vs-Delta-and-g}(c), respectively. One can see that the transmission spectra ($T_{N}$ and $T_{M}$) are symmetric about $\Delta_{2} =2\Gamma$. As $g $ increases, there appear two transmission peaks (valleys) in $T_{N}$ ($T_{M}$). To show more details, we plot $T_{N}$ and $T_{M}$ at different $g$ in Figs.~\ref{Symic-case-TR-vs-Delta-and-g}(b) and \ref{Symic-case-TR-vs-Delta-and-g}(d). When $g/\Gamma=0$, the two giant atoms are decoupled, so the injected photon will have no channel to be transmitted from waveguides M to N. The scattering coefficients $T_{N}=R_{N}=0$ [the red solid curve in Fig.~\ref{Symic-case-TR-vs-Delta-and-g}(b)] and $T_{M}+R_{M}=1$. When $g/\Gamma \in \left( 0,2\right)$, the distance between the two peaks is 0. As a result, $T_{N}$ has one transmission peak locked at $\Delta_{2} =2\Gamma$ [the blue dashed curve for $g/\Gamma=1$ in Fig.~\ref{Symic-case-TR-vs-Delta-and-g}(b)]. When $g/\Gamma >2$,  two transmission peaks appear in $T_{N}$ [the yellow dot-dashed curve for $g/\Gamma=3$ and the green dotted curve for $g/\Gamma=5$ in Fig.~\ref{Symic-case-TR-vs-Delta-and-g}(b)]. For the transmission spectra of $T_{M}$, we find similar features with that of $T_{N}$. However, we point out that although the spectra of $T_{M}$ and $T_{N}$ are symmetric about $\Delta_{2} =2\Gamma$, the locations of the peaks of $T_{N}$ and the valleys of $T_{M}$ are slightly offset. As $g$ increases, the number of the valleys in $T_{N}$ increases from 1 to 2, the distance and depth of the valleys increase accordingly. From the above analysis, we know that the scattering of the single photon can be adjusted by tuning the coupling strength between the two giant atoms.

\begin{figure}[tbp]
	\center\includegraphics[width=0.48\textwidth]{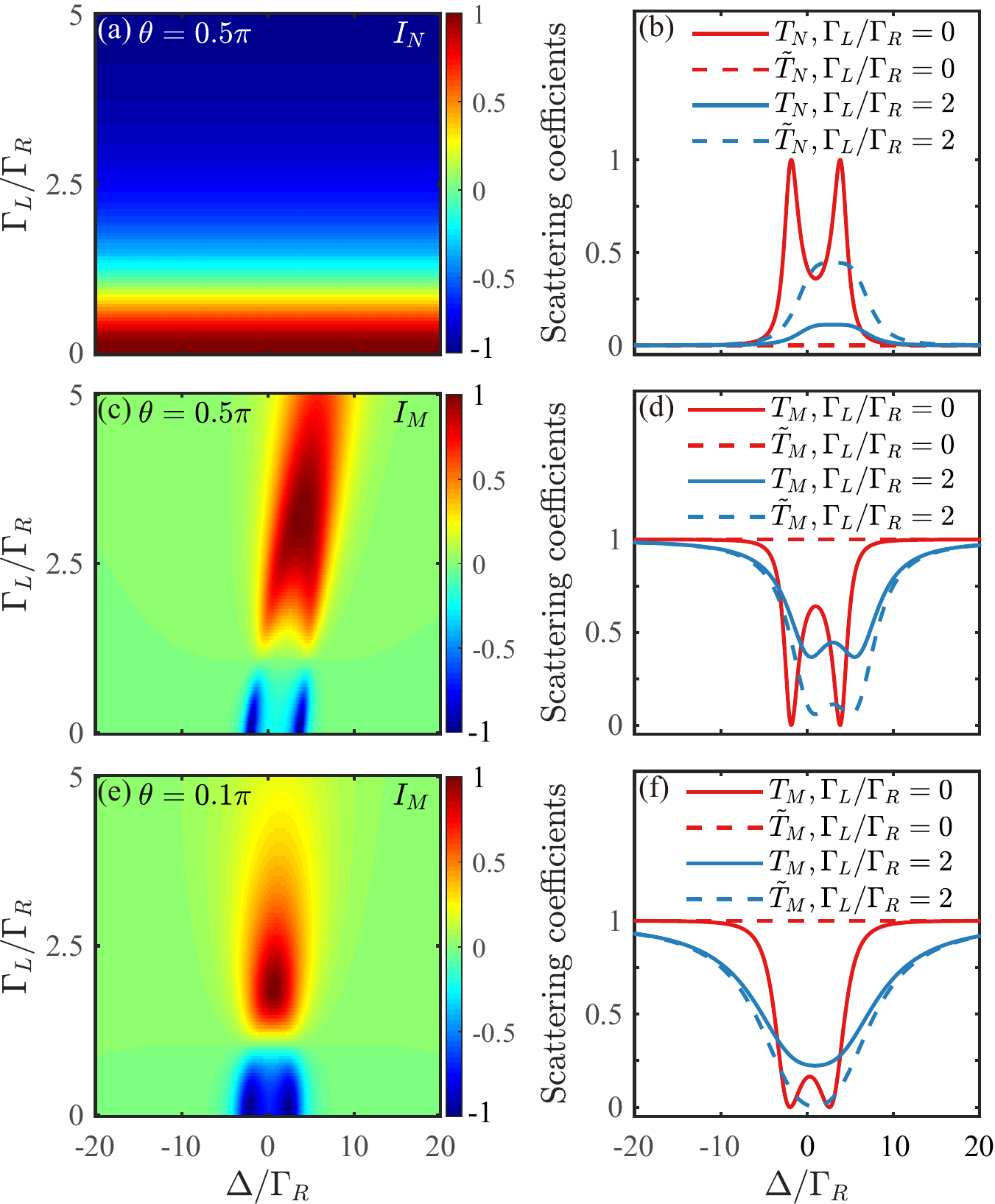}
	\caption{Chiralities (a) $I_{N}$ versus the scaled detuning $%
		\Delta/\Gamma $ and the decay ratio $\Gamma _{L}$/$\Gamma _{R}$ when $\theta _{l}=\theta _{d}=\theta =0.5\pi $. Scattering coefficients (b) $T_{N}$ and $\tilde{T}_{N}$ versus the scaled detuning $\Delta/\Gamma $ at different values of $\Gamma _{L}$/$\Gamma _{R}$.
		$I_{M}$ versus the scaled detuning $%
		\Delta/\Gamma $ and the decay ratio $\Gamma _{L}$/$\Gamma _{R}$ when (c) $\theta =0.5\pi $ and (e) $\theta =0.1\pi $. Scattering coefficients $T_{M}$ and $\tilde{T}_{M}$ versus the scaled detuning $\Delta/\Gamma $ at different values of $\Gamma _{L}$/$\Gamma _{R}$ when (d) $\theta =0.5\pi $ and (f) $\theta =0.1\pi $.
		In all panels, we choose $g/\Gamma =3$.}
	\label{INIM}
\end{figure}

\subsection{Chiral-coupling case}\label{Chiral coupling case}
We now turn to the chiral-coupling case, in which the coupling strengths of the giant atoms with the left-  and right-propagating waveguide modes are different, which leads to asymmetric	scattering, including chiral and nonreciprocal scattering. Below, we first consider the chiral single-photon scattering~\cite{Chen2022}.  To quantitatively describe the chirality in this system, we define the contrast ratio of the scattering coefficients for two opposite directions  (port 1$\rightarrow $port 4 and port 2$\rightarrow $port 3) as

\begin{equation}
	I_{N}=\frac{T_{N}-\tilde{T}_{N}}{T_{N}+\tilde{T}_{N}}=1-\frac{2\Gamma _{\eta
			L}\Gamma _{\lambda L}}{\Gamma _{\eta L}\Gamma _{\lambda L}+\Gamma _{\eta
			R}\Gamma _{\lambda R}},\label{contrast ratioIN}
\end{equation}
where we used Eqs.~(\ref{Sca- amplitude port1 2}) and (\ref{Sca- amplitude port2 2}). It can be seen from Eq.~(\ref{contrast ratioIN}) that this chiral transmission behaviour completely depends on the coupling strength between the giant atoms and the waveguides. Considering a special case, we set $\Gamma _{\eta R}=\Gamma _{\lambda R}=\Gamma _{R}$ and $\Gamma _{\eta L}=\Gamma _{\lambda L}=\Gamma _{L}$.
Figure~\ref{INIM}(a) shows the contrast ratio $I_{N}$ as a decreasing function of the decay ratio $\Gamma _{L}/\Gamma _{R}$. According to Eq.~(\ref{contrast ratioIN}), the contrast ratio $I_{N}=0$ only holds for the symmetric-coupling case, i.e., $\Gamma _{L}$/$\Gamma _{R}=1$, which has been discussed in Sec.~\ref{Symmetric coupling case}. The cases $I_{N}=1$ and $I_{N}=-1$, corresponding to  $\Gamma _{L}$/$\Gamma _{R}=0$ and $\Gamma _{L}$/$\Gamma _{R}\rightarrow \infty $, imply that the chiral scattering reaches its strongest. 

Next we consider the case of the nonreciprocal single-photon scattering.
Similarly, we also introduce contrast ratio $I_{M}$ of the scattering coefficients for two opposite directions (port 1$\rightarrow $port 2 and port 2$ \rightarrow $port 1) to quantitatively describe the nonreciprocity in this system. The expression of $I_{M}$ is given by
\begin{equation}
I_{M}=\frac{T_{M}-\tilde{T}_{M}}{T_{M}+\tilde{T}_{M}}.
\end{equation}

\begin{table}
	\caption{Four ideal chiral-coupling cases in the giant-molecular waveguide-QED system.}
	\label{table1}
	\begin{centering}
		\begin{tabular}{ccccccccc}
			\toprule 
			\multirow{2}{*}{\diagbox{Cases}{Coefficients}} & \multicolumn{4}{c}{port 1 incident} & \multicolumn{4}{c}{port 2 incident}\tabularnewline
			\cmidrule{2-9} \cmidrule{3-9} \cmidrule{4-9} \cmidrule{5-9} \cmidrule{6-9} \cmidrule{7-9} \cmidrule{8-9} \cmidrule{9-9} 
			& $R_{N}$ & $T_{N}$ & $R_{M}$ & $T_{M}$ & $\tilde{R}_{N}$ & $\tilde{T}_{N}$ & $\tilde{R}_{M}$ & $\tilde{T}_{M}$\tabularnewline
			\midrule 
			$\begin{array}{c}
				\Gamma_{\eta R}=\Gamma_{\lambda R}=\Gamma\\
				\Gamma_{\eta L}=\Gamma_{\lambda L}=0
			\end{array}$& 0& $C_{1}$& $0$ & $C_{2}$ & $0$ & 0 & 0 & 1\tabularnewline
			\midrule 
			$\begin{array}{c}
				\Gamma_{\eta R}=\Gamma_{\lambda R}=0\\
				\Gamma_{\eta L}=\Gamma_{\lambda L}=\Gamma
			\end{array}$ & 0 & 0 & 0 & 1 & 0 & $C_{1}$ & 0 & $C_{2}$\tabularnewline
			\midrule 
			$\begin{array}{c}
				\Gamma_{\eta R}=\Gamma_{\lambda L}=\Gamma\\
				\Gamma_{\eta L}=\Gamma_{\lambda R}=0
			\end{array}$ & 0 & 0 & 0 & 1 & $C_{1}$ & 0 & 0 & $C_{2}$\tabularnewline
			\midrule 
			$\begin{array}{c}
				\Gamma_{\eta R}=\Gamma_{\lambda L}=0\\
				\Gamma_{\eta L}=\Gamma_{\lambda R}=\Gamma
			\end{array}$ & $C_{1}$ & 0 & 0 & $C_{2}$ & 0 & 0 & 0 & 1\tabularnewline
			\bottomrule
		\end{tabular}
		\par\end{centering}
\end{table}

In Fig.~\ref{INIM}(c), we plot the contrast ratio $I_{M}$ as a function of the scaled detuning $\Delta/\Gamma $ and the decay ratio $\Gamma _{L}$/$\Gamma _{R}$. In order to illustrate the effects of the chiral-coupling conditions on chiral and asymmetric scattering more clearly, in Figs.~\ref{INIM}(b) and \ref{INIM}(d), we plot the profiles of $T_{N}$ , $\tilde{T}_{N}$,  $T_{M}$, and $\tilde{T}_{M}$ in the ideal chiral-coupling case ($\Gamma _{L}$/$\Gamma _{R}=0$) and the nonideal chiral-coupling case ($\Gamma _{L}$/$\Gamma _{R}=2$), respectively. We can see that, in the ideal chiral-coupling case, when a single photon is injected from port 1, the scattering coefficients $T_{N}$ and $T_{M}$ exhibit double-peak and double-valley line shapes. When a single photon is injected from port 2,  the photon is expected to appear at port 1 with an almost 100\% probability since the giant atom is decoupled from the left-propagating waveguide mode. In the nonideal chiral-coupling case, the single photon can still be asymmetrically scattered, yet the scattering contrast ratio and the asymmetrical effect are degraded in this case. 

Contrast to the chiral scattering described by $I_{N}$, the nonreciprocal scattering that described by $I_{M}$ can also be modulated by the phase shift. As shown in Figs.~\ref{INIM}(c) and \ref{INIM}(e), when the phase shift is changed, the coupling conditions required to achieve a perfect nonreciprocal scattering will change. 
It can be seen that all the $I_{M}$, $T_{M}$, and $\tilde{T}_{M}$ are modulated by the phase shift. Therefore, for the nonreciprocal single-photon scattering that described by $I_{M}$, we can tune both the decay ratio $\Gamma _{L}$/$\Gamma _{R}$ and the phase shift $\theta$ to obtain the most efficient realization of the quantum device. However, for the contrast ratio $I_{N}$, it can only be adjusted by changing the decay ratio $\Gamma _{L}$/$\Gamma _{R}$. 

\begin{figure}[tbp]
	\center\includegraphics[width=0.49\textwidth]{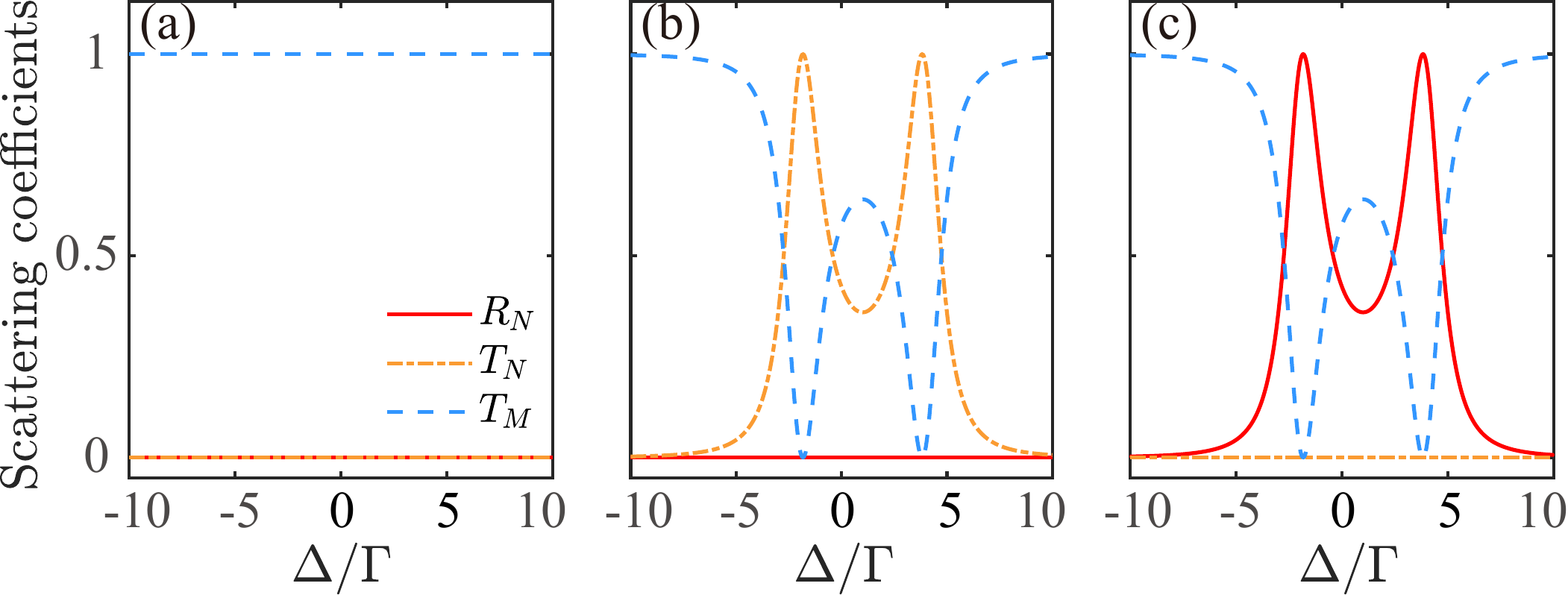}
	\caption{Scattering coefficients $R_{N}$ (red solid curve), $T_{N}$ (yellow dot-dashed curve), and $T_{M}$ (blue dashed curve) versus the scaled detuning $\Delta /\Gamma
		$ for (a) $\Gamma _{\eta R}=\Gamma _{\lambda R}=0$, $\Gamma _{\eta L}=\Gamma
		_{\lambda L}=\Gamma $; (b) $\Gamma _{\eta R}=\Gamma _{\lambda R}=\Gamma $, $%
		\Gamma _{\eta L}=\Gamma _{\lambda L}=0$; (c) $\Gamma _{\eta R}=\Gamma
		_{\lambda L}=0$, $\Gamma _{\eta L}=\Gamma _{\lambda R}=\Gamma $. In all
		panels, we choose $\theta _{l}=\theta _{d}=\theta=0.5\pi $ and $g/\Gamma =3$.}
	\label{Targeted-Router}
\end{figure}

In view of the above discussions, we next focus on the ideal chirality case. To clearly see the influence of chiral coupling on single-photon scattering, we summarize in Table~\ref{table1} the transmission and reflection coefficients of a single photon injected from ports 1 and port 2 in four specific ideal chiral-coupling cases. In Table~\ref{table1}, we introduce
	\begin{subequations}
		\begin{align}
		C_{1} &=\left\vert \frac{i( 1+e^{\pm i\theta _{d}}) (
			1+e^{\pm i\theta _{l}}) g\Gamma }{g^{2}-[ \Delta +i\left(
			1+e^{i\theta _{d}}\right) \Gamma ] [ \Delta +i\left( 1+e^{i\theta
				_{l}}\right) \Gamma ] }\right\vert ^{2}, \\
		C_{2} &=\left\vert \frac{g^{2}-[ \Delta +i( 1+e^{i\theta
				_{d}}) \Gamma ] [ \Delta -i( 1+e^{-i\theta
				_{l}}) \Gamma ] }{g^{2}-[ \Delta +i\left( 1+e^{i\theta
				_{d}}\right) \Gamma ] [ \Delta +i\left( 1+e^{i\theta _{l}}\right)
			\Gamma ] }\right\vert ^{2}.
			\end{align}
	\end{subequations}
We find that in the four chiral-coupling cases summarized in Table~\ref{table1}, the giant molecule can exhibit chiral and nonreciprocal
transmissions. It can be proved that $C_{1}$ and $C_{2}$ satisfy the relation $C_{1}+C_{2}=1$, and hence we can make $C_{1}=1, C_{2}=0$ or $C_{1}=0, C_{2}=1$ [see the yellow dot-dashed curve and the red solid curve in Fig.~\ref{Targeted-Router}] by selecting some special parameters. This means that, perfect chiral and nonreciprocal scattering can be achieved in the ideal chiral-coupling case. When a single photon is injected from a certain port of waveguide M, the probability for detecting the photon at a specified port (the other three ports) can reach 100\%, by adjusting the coupling condition of the system. This issue has received a lot of attention, and we will focus on it in the next subsection.

\subsection{Targeted routing in the Markovian regime}
Based on several ideal chiral-coupling cases summarized in Table~\ref{table1}, in this section, we demonstrate how to achieve directional routing of a single photon in this system. That is, when a photon is injected from one port, we can route it to other three ports on demand. 

We plot in Fig.~\ref{Targeted-Router} the profiles of $T_{M}$ (the blue dashed curve), $T_{N}$ (the yellow dot-dashed curve), and $R_{N}$ (the red solid curve) versus the scaled detuning $\Delta/\Gamma $ for different ideal chiral-coupling cases. As shown in Fig.~\ref{Targeted-Router}(a),
when $\Gamma _{\eta R}=\Gamma _{\lambda R}=0$ and $\Gamma _{\eta L}=\Gamma_{\lambda L}=\Gamma $, we have $T_{M}=1, T_{N}=0$, and $R_{N}=0$, which means that the single photon can only be routed to port 2 of waveguide M. As shown in Fig.~\ref{Targeted-Router}(b), when $\Gamma _{\eta R}=\Gamma _{\lambda R}=\Gamma $ and $\Gamma _{\eta L}=\Gamma_{\lambda L}=0$, the scattering coefficient $T_{N}$ has two peaks located at $\Delta =\Gamma \sin \theta \pm \sqrt{g^{2}-3\Gamma^{2}/2-2\Gamma ^{2}\cos \theta -\Gamma ^{2}\cos \left( 2\theta \right) /2}$. In this case, we have $T_{M}=0, T_{N}=1$, and $R_{N}=0$. Namely, the single photon can only be routed to port 4 of waveguide N. As shown in Fig.~\ref{Targeted-Router}(c), for $\Gamma _{\eta R}=\Gamma _{\lambda L}=0$ and $\Gamma _{\eta L}=\Gamma_{\lambda R}=\Gamma $, we know that the scattering coefficient $R_{N}$ with two peaks located at $\Delta =\Gamma \sin \theta \pm \sqrt{g^{2}-3\Gamma^{2}/2-2\Gamma ^{2}\cos  \theta -\Gamma ^{2}\cos \left( 2\theta \right) /2}$, where $T_{M}=0, T_{N}=0$, and $R_{N}=1$. Hence the photon can be routed to port 3 of waveguide N totally. The above discussions demonstrate that it is possible to realize directional single-photon routing. Our system is a good platform for realizing a single-photon router.

\section{Single photon scattering in the non-Markovian regime}\label{Non-Markovian regime}
In the previous discussions, we have analyzed the chiral and nonreciprocal single-photon scattering by neglecting the propagating time of photons in the waveguides. For the giant molecule with multiple coupling points, however, the investigation of the influence of the non-Markovian retarded effect on the scattering behaviors is also interesting. When the propagating time $\tau _{l}$ ($\tau _{d}$) is comparable to or larger than the atomic lifetime, i.e., $\{ \tau _{l}\Gamma _{\lambda R},\tau _{l}\Gamma _{\lambda L},\tau
	_{d}\Gamma _{\eta R},\tau _{d}\Gamma _{\eta L}\} \approx 1$ or $\{ \tau _{l}\Gamma _{\lambda R},\tau _{l}\Gamma _{\lambda L},\tau
	_{d}\Gamma _{\eta R},\tau _{d}\Gamma _{\eta L}\} >1$, the non-Markovian retarded effect induced by $\tau _{l}$ ($\tau _{d}$) cannot be neglected. In this case, the giant atoms enter the non-Markovian regime and then the phases $\varphi _{l}=\tau _{l}\Delta +\theta _{l}$ and $\varphi _{d}=\tau _{d}\Delta +\theta _{d}$ are both sensitive to the detuning  $\Delta $. Hence we can expect the transmission spectra to be more complex. In realistic systems, when a transmon qubit couples with surface acoustic waves or when the distance between the coupling points is large enough, the non-Markovian effect should be taken into account~\cite{Andersson2019,Guo2017}. 

\begin{figure}[tbp]
	\center\includegraphics[width=0.48\textwidth]{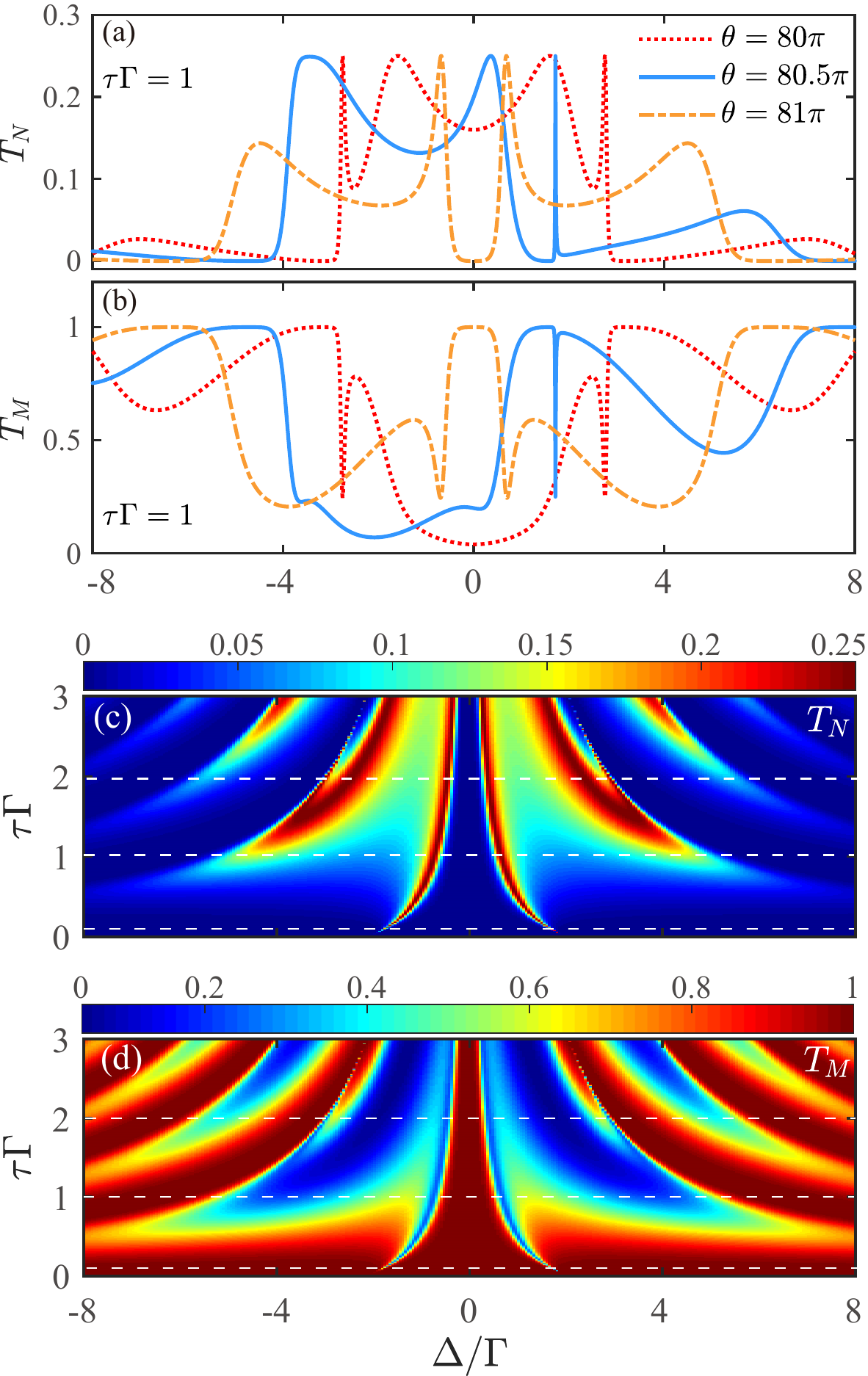}
	\caption{Transmission coefficients (a) $T_{N}$ and (b) $T_{M}$ as functions of the scaled detuning $\Delta/\Gamma $ at $\theta=80\pi $ (the red dotted curve), $\theta=80.5\pi $ (the blue solid curve), and $\theta=81\pi $ (the yellow dot-dashed curve). The other parameter in panels (a) and (b) is $\tau \Gamma =1$. Transmission coefficients (c) $T_{N}$ and (d) $T_{M}$ as functions of $\Delta/\Gamma $ and $\tau\Gamma$ with $\theta =81\pi $. In all panels, we consider the symmetric-coupling case, i.e., $\Gamma_{L}=\Gamma_{R}=\Gamma$. In all panels, we choose $g/\Gamma =2$.}\label{SymmtricNonMarkovian}
\end{figure}

\subsection{Symmetric- and Chiral-coupling cases}
We first focus on the symmetric-coupling case and $\varphi _{l}=\varphi _{d}=\varphi =\tau\Delta +\theta$. Figure~\ref{SymmtricNonMarkovian}(a) [\ref{SymmtricNonMarkovian}(b)] shows the transmission coefficient $T_{N}$ ($T_{M}$) as a function of the scaled detuning $\Delta/\Gamma$ at phases $\theta =80\pi$, $\theta =80.5\pi$, and $\theta =81\pi $. When $\theta =n\pi $ with $n$ an integer, both the transmission spectra $T_{N}$ and $T_{M}$ are symmetric to $\Delta =0$ due to the relation $T_{N/M}( \Delta ,\theta) =T_{N/M}( -\Delta ,\theta )$. When $\theta \neq n\pi $, the transmission spectra become asymmetric. In both cases, the transmission coefficients are characterized by the complicated line shapes with staggered peaks (valleys) and multiple dips. In particular, when the phase shift satisfies $\varphi=\tau\Delta +\theta=(2m+1)\pi$ with $m$ an integer, we can see that many transmission dips with $T_{N}=0$ appear. This is because at $\varphi=(2m+1)\pi$, the giant atoms are decoupled from the waveguides, and hence the incident photon cannot be transmitted to waveguide N. The moving dips in these three cases arise from the fact that $\Delta $ satisfies the relation $\varphi=\tau\Delta +\theta=(2m+1)\pi$ and the value of $\Delta $ varies with the change of $\varphi$. These characteristics are quite different from those in the Markovian regime.

\begin{figure}[tbp]
	\center\includegraphics[width=0.48\textwidth]{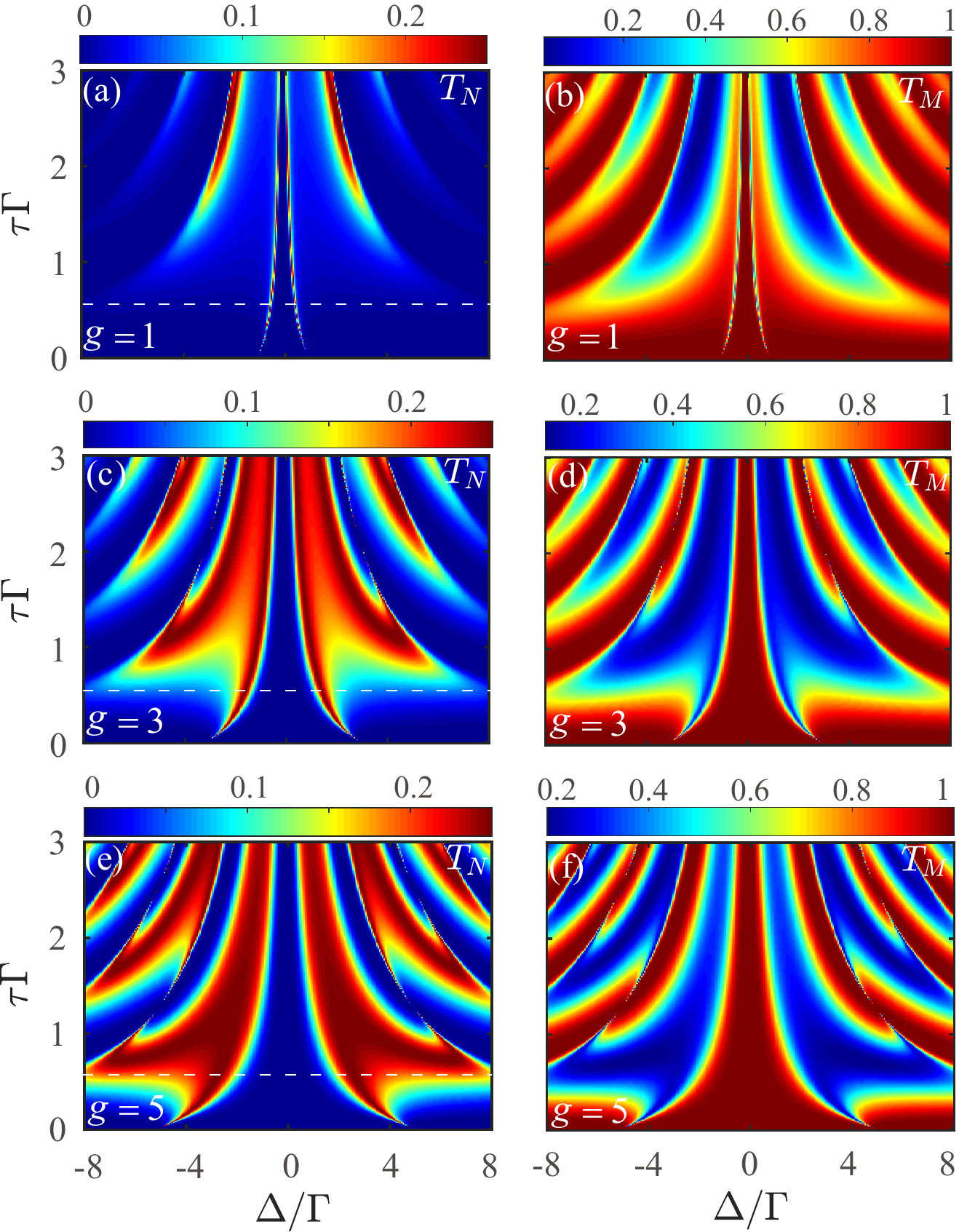}
	\caption{Scattering coefficients $T_{N}$ and $T_{M}$ versus the scaled detuning $\Delta/\Gamma$ and the scaled propagating time $\tau\Gamma$ at different coupling strengths. The left and right columns represent $T_{N}$ and $T_{M}$, respectively. The first, second, and third rows correspond to $g/\Gamma=1$,  $3$, and $5$, respectively. In all panels, we choose $\Gamma_{L}=\Gamma_{R}=\Gamma$ and $\theta _{l}=\theta _{d}=\theta =81\pi $. }
	\label{g-In-NonMarkovian}
\end{figure}

\begin{figure}[tbp]
	\center\includegraphics[width=0.48\textwidth]{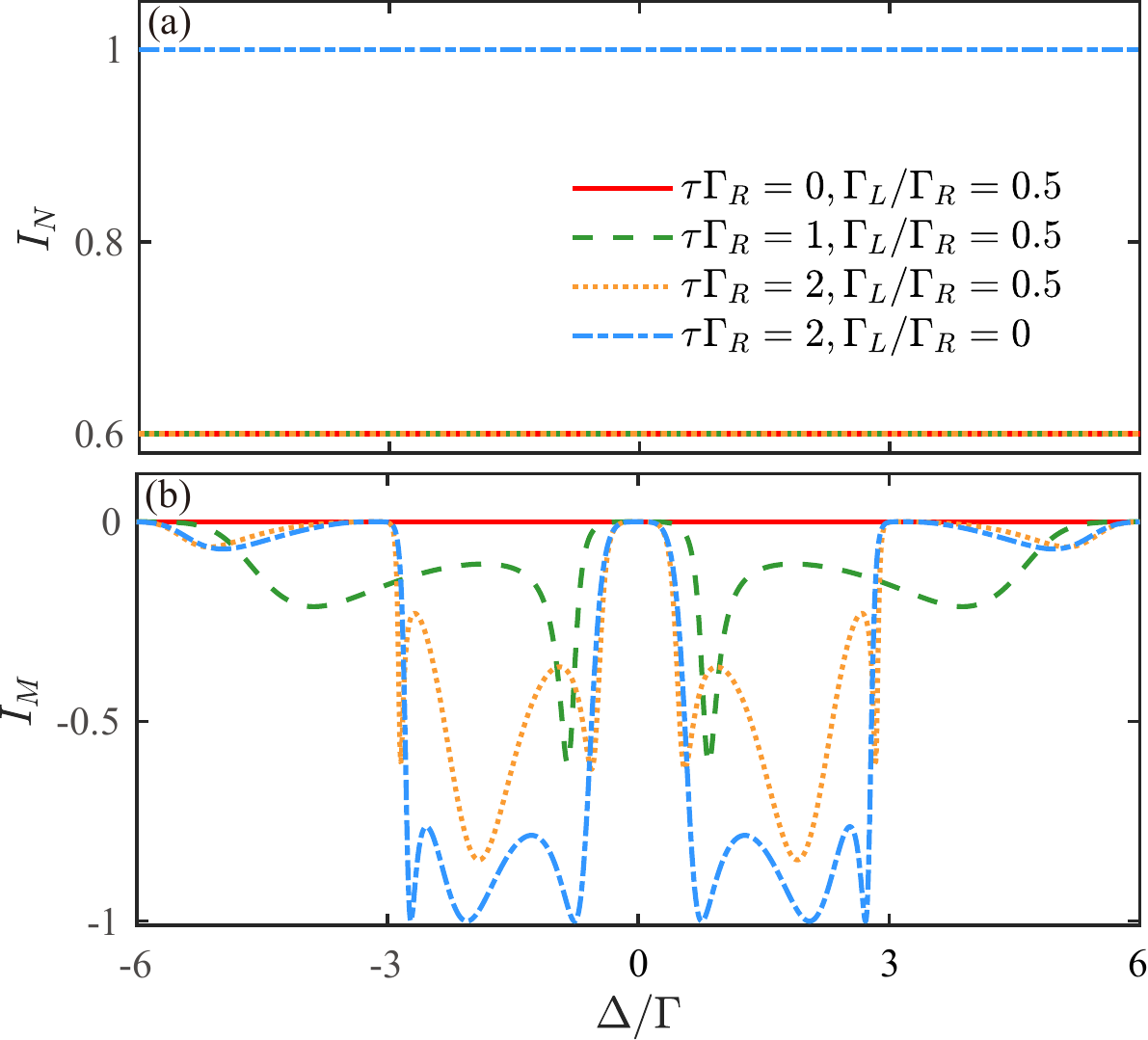}
	\caption{Transmission contrast ratios (a) $I_{N}$ and (b) $I_{M}$ as functions of $\Delta/\Gamma$ with different values of $\tau\Gamma _{R}$ and $\Gamma _{L}/\Gamma _{R}$ in the cases of $\tau\Gamma _{R}=0$ and $\Gamma _{L}/\Gamma _{R}=0.5$ (the red solid curve),  $\tau\Gamma _{R}=1$ and $\Gamma _{L}/\Gamma _{R}=0.5$ (the green dashed curve), $\tau\Gamma _{R}=2$ and $\Gamma _{L}/\Gamma _{R}=0.5$ (the yellow dotted curve), and $\tau\Gamma _{R}=2$ and $\Gamma _{L}/\Gamma _{R}=0$ (the blue dot-dashed curve). In all panels, we assume $\theta_{l}=\theta _{d}=\theta=81\pi$ and $g/\Gamma =2$.}\label{ChiralNonMarkovian}
\end{figure}

We point out that, there is no determined boundary between the Markovian and non-Markovian regimes. The behavior of the single-photon scattering changes continuously when the system transits from the Markovian regime to non-Markovian regime. To better see this transition process, we plot the transmission coefficient $T_{N}$ ($T_{M}$) as a function of $\Delta/\Gamma $ and $\tau\Gamma$ in Fig.~\ref{SymmtricNonMarkovian}(c) [\ref{SymmtricNonMarkovian}(d)] when the coupling is symmetric. It shows a significant difference between the two regimes. In the Markovian regime, the single photon is completely routed to port 2 when $\theta=n\pi$. However, as the propagating time $\tau\Gamma$ increases from zero to a larger value, the system enters the non-Markovian regime. It can be seen that the spectra exhibit continuous changing process. In particular, the spectra are characterized by a more abundant and multiple-peak structure due to the stronger non-Markovian retarded effect. Comparing the profiles of $T_{N}$ ($T_{M}$) at $\tau\Gamma=0.1$, $1$, and $2$ [see the while dashed lines in Figs.~\ref{SymmtricNonMarkovian}(c)  and~\ref{SymmtricNonMarkovian}(d)], the spectra are more sensitive to the change of  the scaled detuning $\Delta$.

Below, we study the effect of the coupling strength between the two giant atoms on the single-photon scattering during the transition of the system from the Markovian to non-Markovian regimes. In Fig.~\ref{g-In-NonMarkovian}, we plot the transmission coefficients $T_{N}$ and $T_{M}$ as functions of the scaled detuning $\Delta/\Gamma $ and the scaled propagating time $\tau\Gamma$ when $g/\Gamma=1$,  $3$, and  $5$, respectively. It can be seen that the locations of the peaks and the width of the scattering spectra become different. As $g/\Gamma$ increases, the parameter range of implementing quantum routing of the single photon from waveguides M to N is widened [see Figs.~\ref{g-In-NonMarkovian}(a),~\ref{g-In-NonMarkovian}(c), and~\ref{g-In-NonMarkovian}(e)]. For example, we can take a small value of propagating time $\tau\Gamma=0.5$ to realize the quantum routing in a wide range of $\Delta/\Gamma $, as shown by the white dashed lines in Figs.~\ref{g-In-NonMarkovian}(a),~\ref{g-In-NonMarkovian}(c), and~\ref{g-In-NonMarkovian}(e). For the scattering coefficient $T_{M}$, however,
	the transmission of the single photon from ports 1 to 2 in waveguide M is suppressed in a wide range of $\Delta/\Gamma $ with the increase of $g/\Gamma$ [see Figs.~\ref{g-In-NonMarkovian}(b),~\ref{g-In-NonMarkovian}(d), and~\ref{g-In-NonMarkovian}(f)]. The above analyses suggest that the inner coupling strength of the two giant atoms also plays an important role in the single-photon scattering when the system transits from the Markovian to non-Makrovian regimes.

 We now turn to the chiral and nonreciprocal scattering in both the Markovian and non-Markovian regimes. In Fig.~\ref{ChiralNonMarkovian}(a) [\ref{ChiralNonMarkovian}(b)] we plot the transmission contrast ratio $I_{N}$ ($I_{M}$) as a function of the scaled detuning $\Delta/\Gamma$. For a given decay ratio $\Gamma _{L}/\Gamma _{R}=0.5$, the value of $I_{N}$ is independent of the propagation time. This feature can be seen from Eq.~(\ref{contrast ratioIN}), which is valid in both two regimes [see the red solid, green dashed, and yellow dotted curves in Fig.~\ref{ChiralNonMarkovian}(a)]. For the contrast ratio $I_{M}$, as shown the red solid curve in Fig.~\ref{ChiralNonMarkovian}(b), the transmission behavior is always reciprocal in the Markovian limit, even when the giant molecule is chirally coupled to the waveguides. However, in the non-Markovian limit, the $I_{M}$ is characterized by more complex line shapes. As the propagating time $\tau\Gamma$ increases from zero to a larger value, the maximal value of $|I_{M}|$ can be gradually enhanced until it approaches one. Therefore, the non-Markovian retarded effect induced by $\tau\Gamma$ provides us an extra choice to improve the nonreciprocity when the system is in the non-Markovian regime [see the green dashed and yellow dotted curves in Fig.~\ref{ChiralNonMarkovian}(b)]. Note that the chiral/nonreciprocal scattering is not perfect at $\Gamma _{L}/\Gamma _{R}=0.5$ ($\vert I_{N/M}<1\vert $). To realize a perfect chiral/nonreciprocal scattering in the non-Markovian regimes, we also need to consider the ideal chiral-coupling case, as shown the blue dot-dashed curve in Fig.~\ref{ChiralNonMarkovian}. 

\subsection{Targeted routing in the non-Markovian regime}
\begin{figure}[tbp]
	\center\includegraphics[width=0.48\textwidth]{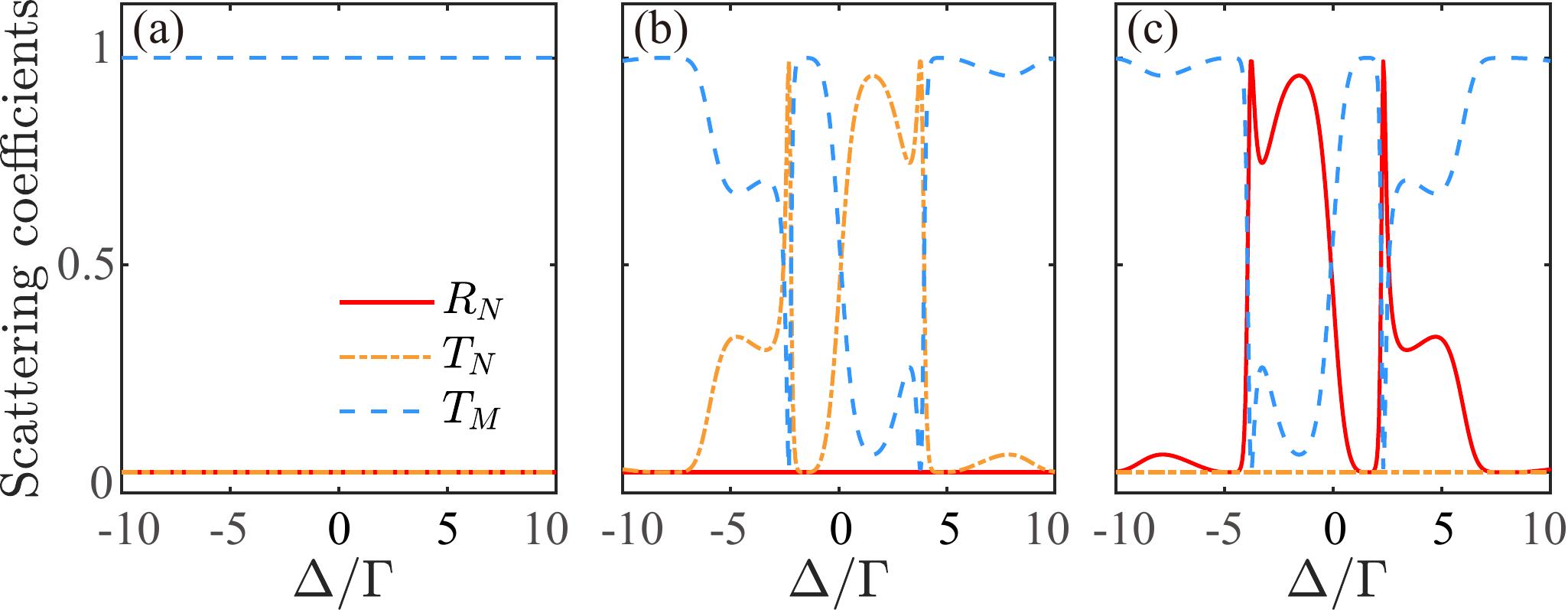}
	\caption{Scattering coefficients $R_{N}$ (red solid curve), $T_{N}$ (yellow dot-dashed curve), and $T_{M}$ (blue dashed curve) versus the scaled detuning $\Delta /\Gamma
		$ for (a) $\Gamma _{\eta R}=\Gamma _{\lambda R}=0$, $\Gamma _{\eta L}=\Gamma
		_{\lambda L}=\Gamma $; (b) $\Gamma _{\eta R}=\Gamma _{\lambda R}=\Gamma $, $%
		\Gamma _{\eta L}=\Gamma _{\lambda L}=0$; (c) $\Gamma _{\eta R}=\Gamma
		_{\lambda L}=0$, $\Gamma _{\eta L}=\Gamma _{\lambda R}=\Gamma $. In all
		panels, we choose $\theta _{l}=\theta _{d}=\theta=80.5\pi $, $g/\Gamma =3$, and $\tau\Gamma=1$.}\label{TargetedRouterInNonmarkovian}
\end{figure}

In the following, we show that the single-photon directional routing can also be realized when the system is in the non-Markovian regime. Figure~\ref{TargetedRouterInNonmarkovian} shows the scattering coefficients from port 1 to other three ports versus the scaled detuning $\Delta /\Gamma$. Here, we choose $\theta _{l}=\theta _{d}=\theta=80.5\pi$ and $\tau\Gamma=1$. It can be seen that, by adjusting the coupling strengths of the giant molecule to the left- and right-propagating fields in the waveguides, the single photon injected from port $1$ can be sent to ports $2$, $3$, and $4$ on demand, respectively. For the parameters $\Gamma _{\eta R}=\Gamma _{\lambda R}=0$ and $\Gamma _{\eta L}=\Gamma
_{\lambda L}=\Gamma $, as shown in Fig.~\ref{TargetedRouterInNonmarkovian}(a), the scattering spectra have the same line shapes with those in the Markovian regime [see in Fig.~\ref{Targeted-Router}(a)]. However, when we take $\Gamma _{\eta R}=\Gamma _{\lambda R}=\Gamma $ and $\Gamma _{\eta L}=\Gamma _{\lambda L}=0$ [$\Gamma _{\eta R}=\Gamma_{\lambda L}=0$ and $\Gamma _{\eta L}=\Gamma _{\lambda R}=\Gamma $], as shown in Fig.~\ref{TargetedRouterInNonmarkovian}(b) [Fig.~\ref{TargetedRouterInNonmarkovian}(c)], the scattering spectra are asymmetric to $\Delta/\Gamma=0$ and exhibit more complicated line shapes. In particular, the single photon can not only be sent to port $2$ but also to port $3$ ($4$), as shown by the blue dashed curve in Fig.~\ref{TargetedRouterInNonmarkovian}(b) [Fig.~\ref{TargetedRouterInNonmarkovian}(c)], which are different from the cases of the Markovian limit. Since this phenomenon arises from the non-Markovian retarded effect, we would like to refer to it as non-Markovian-induced directional single photon routing.

\section{Discussions on the experimental implementation of this scheme}\label{Experimental}

In this section, we present some discussions on the possible experimental implementation of the chiral-giant-molecule waveguide-QED system. The key element for the experimental implementation of this device is to realize the chiral coupling between the giant atoms with the waveguides at each coupling point and the inner coupling between the two giant atoms.  In the case of small atoms chirally coupled to a 1D waveguide, the chiral system-reservoir interaction has been realized in various experimental platforms  by using circularly polarized light modes, such as a gold nanoparticle~\cite{Petersen2014} or cesium atoms~\cite{Mitsch2014} interacting with a tapered nanofiber, and semiconducting quantum dots coupled to photonic crystal waveguides~\cite{sollner2015}. In addition, it can be realized in other platforms, such as phononic waveguides made of Bose-Einstein quasicondensates~\cite{Lin2011,Galitski2013,Ramos2014}, magnonic waveguides consisting of spin chains~\cite{Vermersch2016,Ramos2016} or yttrium iron garnet~\cite{Loo2018}. However, the implementation of the chiral coupling between the giant atom and the waveguide at multiple coupling points is not a trivial task~\cite{Kockum2021}. Though some theoretical works have studied the chiral coupling of the giant atoms between the waveguide modes~\cite{Carollo2020,Soro2022,Xin Wang2022}, to our best knowledge, there is no experimental demonstration of the chiral coupling in our proposal model. Recently, some proposals for the possible realization of the chiral couplings have been proposed, for example, utilizing superconducting qubits coupled to a transmission line with circulators inserted to provide the chirality~\cite{Gu2017}. Nevertheless, the detailed investigation concerning the experimental implementation of the chiral coupling between the giant atoms and waveguides is not sufficient, and it deserves to be studied in specific works.

To realize the coupling between the giant molecule with the waveguide at four points, we can use two Xmon qubits with inner coupling strength to couple to two individual meandering transmission lines~\cite{Kannan2020}.  Note that the inner coupling between the two qubits can be implemented by coupling them to a tunable inductor, with accessible coupling strength up to tens of MHz~\cite{Chen2014}. Each qubit interacts with its transmission line at two well separated locations. The distance $l$ ($d$) between the two coupling points of the qubits $a$ ($b$) can be set precisely by changing the transmission line length~\cite{Kockum2014}.  In the Markovian regime, photons in the waveguide M (N) will experience a phase shift $\varphi _{l}\approx\theta_{l}=\omega _{0}l/\upsilon _{g}=2\pi l/\lambda \left( \omega
\right) $ [$\varphi _{d}\approx\theta_{d}=\omega _{0}d/\upsilon _{g}=2\pi d/\lambda \left( \omega
\right) $], which depends on the distance between the coupling points and the wavelength $\lambda \left( \omega \right) $ for modes at frequency $\omega =\omega _{0}$ of the two Xmon qubits. Experimentally, the phase shift $\varphi _{l}$ ($\varphi _{d}$) can be adjusted by changing the frequencies of the qubits, and thereby the wavelength $\lambda \left( \omega \right) =2\pi \upsilon _{g}/\omega $. In addition, we can obtain the corresponding relation between the typical distance and wavelength when the phase shift is fixed. For example, if we take $\theta _{l}=\theta _{d}=0.5\pi $, the distances between the coupling points can be obtained as $l=d=\lambda(\omega) /4$. In the non-Markovian regime, the phase shift includes a $\Delta$-dependent term and then becomes $\varphi _{l}=\tau _{l}\Delta +\theta _{l}$ ($\varphi _{d}=\tau _{d}\Delta +\theta _{d}$), with the propagating time of the photons in the waveguide being $\tau _{l}=l/\upsilon _{g}$ ($\tau _{d}=d/\upsilon _{g}$). To work in the non-Markovian regime, we can increase the distances between two coupling points to meet $\tau _{l}\approx 1$ ($\tau _{d}\approx 1$)~\cite{Kannan2020}, and hence the single-photon scattering in the non-Markovian regime can be realized.

\section{Conclusion}\label{Conclusions}
In conclusion, we have studied the single-photon scattering in a chiral-giant-molecule waveguide-QED system under the Markovian and non-Markovian limits, in which the phase shifts are detuning-independent and detuning-dependent, respectively. We have found that it is possible to control the single-photon scattering behavior by tuning the phase shifts between the coupling points, the coupling strength between the two giant atoms, and the coupling strength between the giant molecule and the waveguides. In the Markovian regime, we have studied the single-photon scattering in both the symmetric-coupling and chiral-coupling cases. In the symmetric-coupling case, the single-photon scattering is achiral and reciprocal, whereas, in the chiral-coupling case,  chiral and nonreciprocal single-photon scattering is achieved. To realize perfect  chirality and nonreciprocity, we have further investigated the single-photon scattering in the ideal chiral-coupling case, where the single-photon directional router is realized. In the non-Markovian regime, the scattering spectra with multiple peaks and staggered dips become more abundant.  These results are expected to be applied in quantum information processing and quantum device design at the single-photon level. This work will pave the way to the study of single-photon quantum devices in giant-molecular waveguide-QED systems. In the future research, it is an interesting topic to study the chiral coupling of the giant atom or molecule with the topological waveguides, which could be regarded as a good platform to analyze the topological properties of the fields coupled to the giant atoms, such as the relation between the positions of the scattering matrix poles with the non-trivial Chern numbers.

\begin{acknowledgments}
We would like to thank Xun-Wei Xu and Xin Wang for helpful discussions. J.-Q.L. was supported in part by National Natural Science Foundation of China (Grants No. 12175061, No. 12247105, and No. 11935006), the Science and Technology Innovation Program of Hunan Province (Grants No. 2021RC4029 and No. 2020RC4047).
\end{acknowledgments}

\end{document}